\documentclass[fleqn,usenatbib]{mnras}

\usepackage[T1]{fontenc}

\DeclareRobustCommand{\VAN}[3]{#2}
\let\VANthebibliography\thebibliography
\def\thebibliography{\DeclareRobustCommand{\VAN}[3]{##3}\VANthebibliography}

\usepackage{amsmath,amstext,physics}
\usepackage{float}
\usepackage[export]{adjustbox}
\usepackage{graphicx}
\usepackage{printlen,enumitem}
\usepackage{txfonts} 
\usepackage[figure,figure*]{hypcap} 
\usepackage{sidecap}
\usepackage{comment}
\usepackage{subcaption}
\usepackage{xcolor}
\usepackage[normalem]{ulem}

\newcommand{\simi}{{\bf \texttt{m12i}}}
\newcommand{\simb}{{\bf \texttt{m12b}}}
\newcommand{\simf}{{\bf \texttt{m12f}}}
\newcommand{\simm}{{\bf \texttt{m12m}}}

\newcommand\msun[1]{{\rm M}_{\odot}}
\newcommand{\Msun}{{\rm M}_{\odot}}

\newcommand{\Mdot}{{\dot M}}

\newcommand{\Mbh}{M_{\rm BH}}
\newcommand{\Mdisk}{M_{\rm AD}}

\newcommand{\MdotBH}{\Mdot_{\rm BH}}

\title[Seyfert galaxies on FIRE]{Investigating black hole accretion and feedback self-regulation in Seyfert galaxies using the FIRE-3 cosmological hydrodynamic simulations}

\author[J. Mercedes-Feliz et al.]{Jonathan~Mercedes-Feliz,$^{1}$\thanks{E-mail: jonathan.mercedes\_feliz@uconn.edu}
Daniel Angl{\'e}s-Alc{\'a}zar,$^{1}$
Jose Cevallos,$^{1}$
Boon Kiat Oh,$^{2,1}$\newauthor
Santiago Garc{\'i}a-Burillo,$^{3}$
Rachel K. Cochrane,$^{4}$
Cristina Ramos Almeida,$^{5,6}$\newauthor
Claude-Andr{\'e} Faucher-Gigu{\`e}re,$^{7}$
Almudena Alonso-Herrero,$^{8}$
Alexander J. Richings,$^{9,10}$\newauthor
Miguel Pereira-Santaella,$^{11}$
Jorge Moreno,$^{12}$
Niranjan~Chandra~Roy,$^{1}$
Tanio D{\'i}az-Santos,$^{13,14}$\newauthor
and Philip F. Hopkins$^{15}$
\\
$^{1}$Department of Physics, University of Connecticut, 196 Auditorium Road, U-3046, Storrs, CT 06269-3046, USA\\
$^{2}$School of Physics, Korea Institute for Advanced Study, 85 Hoegiro, Dongdaemun-gu, Seoul 02455, Republic of Korea\\
$^{3}$Observatorio Astron{\'o}mico Nacional (OAN-IGN)-Observatorio de Madrid, Alfonso XII, 3, 28014-Madrid, Spain\\
$^{4}$Jodrell Bank Centre for Astrophysics, University of Manchester, Oxford Road, Manchester M13 9PL, UK\\
$^{5}$Instituto de Astrof{\'i}sica de Canarias, Calle V{\'i}a L{\'a}ctea, s/n, E-38205, La Laguna, Tenerife, Spain\\
$^{6}$Departamento de Astrof{\'i}sica, Universidad de La Laguna, E-38206, La Laguna, Tenerife, Spain\\
$^{7}$CIERA and Department of Physics and Astronomy, Northwestern University, 1800 Sherman Ave., Evanston, IL 60201, USA\\
$^{8}$Centro de Astrobiolog{\'i}a (CAB), CSIC-INTA, Camino Bajo del Castillo s/n, E-28692 Villanueva de la Ca\~nada, Madrid, Spain\\
$^{9}$Centre for Data Science, Artificial Intelligence and Modelling, University of Hull, Cottingham Road, Hull, HU6 7RX, UK\\
$^{10}$E. A. Milne Centre for Astrophysics, University of Hull, Cottingham Road, Hull, HU6 7RX, UK\\
$^{11}$Instituto de F{\'i}sica Fundamental, CSIC, Calle Serrano 123, 28006 Madrid, Spain\\
$^{12}$Department of Physics and Astronomy, Pomona College, 333 N. College Way, Claremont, CA 91711, USA\\
$^{13}$Institute of Astrophysics, Foundation for Research and Technology–Hellas (FORTH), Heraklion 70013, Greece\\
$^{14}$School of Sciences, European University Cyprus, Diogenes Street, Engomi 1516, Nicosia, Cyprus\\
$^{15}$TAPIR, Mailcode 350-17, California Institute of Technology, Pasadena, CA 91125, USA
}

\date{Accepted XXX. Received YYY; in original form ZZZ}

\pubyear{2026}

\begin{document}
\label{firstpage}
\pagerange{\pageref{firstpage}--\pageref{lastpage}}
\maketitle

\begin{abstract}
Observations of local Seyfert galaxies show an intriguing connection between AGN luminosity and a deficit of molecular gas on $\sim$50\,pc scales compared to 200\,pc, the plausible imprint of AGN feedback.
Motivated by these findings, we investigate the interplay between supermassive black hole (BH) accretion, AGN feedback, and nuclear gas reservoirs using cosmological hydrodynamic simulations implementing FIRE-3 multi-phase interstellar medium (ISM) physics and multi-component BH accretion and feedback models. Focusing on the late-time evolution of four Milky Way-mass galaxies, we find recurrent cycles of increased gas inflow toward the accretion disc, enhanced BH accretion, feedback self-regulation, and suppressed gas inflow rate until the next fuelling event. 
AGN winds interact with the ISM and escape preferentially through low-density polar channels after opening central cavities on $\sim$10--500\,pc scales, regulating BH growth and producing episodic behaviour on $\sim$10--100\,Myr timescales.
The simulations reproduce the observed diversity of nuclear morphologies, gas concentrations, and AGN luminosities in late-type Seyfert galaxies, but do not exhibit a clear anti-correlation between gas concentration and AGN luminosity.
Higher-luminosity AGN ($L_{\rm X} \sim 10^{41.5-43}\,{\rm erg\,s}^{-1}$) powered by the accretion disc reservoir can coexist with feedback-driven cavities, consistent with observations, but they are more common in simulated galaxies with centrally-peaked gas distributions.
Although differences in sample selection, tracer choice, spatial resolution, and stochasticity in AGN fuelling may impact underlying concentration–luminosity trends, the apparent tension between simulations and observations points to the timing between gas inflow, accretion-disc depletion, and feedback-driven clearing on $\sim$50--200\,pc scales as a key constraint on AGN self-regulation models.
\end{abstract}




\section{Introduction} \label{sec:intro}
The co-evolution of galaxies and their central supermassive black holes (SMBHs) inferred from observations is thought to be governed by a complex interplay between gas inflow and feedback processes \citep{Heckman&Best2014,Alexander2025}. Accretion of gas onto SMBHs powers active galactic nuclei (AGN), whose radiative and mechanical feedback can significantly impact their host galaxies by heating, dispersing, or expelling gas across scales \citep{Garcia-Burillo2014,Hlavacek-Larrondo2015, harrison2018,DiMatteo2023,Harrison2024}. Observational and theoretical efforts increasingly suggest that this relationship may be self-regulating, wherein phases of efficient gas inflow trigger luminous AGN activity, which in turn limits further accretion by altering the local interstellar medium (ISM).

This concept of self-regulated growth is consistent with a number of empirical correlations observed in the local universe \citep{Magorrian1998,Kormendy&Ho2013}, although these scaling relations do not by themselves establish AGN feedback as the causal mechanism. Tight scaling relations between SMBH mass and host-galaxy properties, such as the stellar velocity dispersion  \citep{Ferrarese2000,Gebhardt2000} and the bulge and total stellar mass of galaxies  \citep{Haring2004,Reines2015,Savorgnan2016,Greene2020,Zeltyn2025}, suggest a long-term co-evolutionary connection. Moreover, observations of multi-phase AGN-driven outflows spanning ionized to molecular gas from the local universe \citep{Garcia-Burillo2014,RamosAlmeida2019,RamosAlmeida2022,Wylezalek2020,Juneau2022,AlonsoHerrero2023} to cosmic noon \citep{Zakamska2016,Circosta2018,Kakkad2020} and the high redshift universe  \citep{Bischetti2024,Onoue2025,Vayner2025} provide direct evidence that AGN can couple energetically to galaxy-scale reservoirs \citep{cicone2014,Fiore2017,harrison2018,Bertola2024}. 

Analytical models \citep[e.g.,][]{Silk1998,Murray2005,Shankar2020} and hydrodynamic simulations \citep{DiMatteo2005,Hopkins2007_bhplane,Choi2012,Dubois2012,Habouzit2017} indeed show that AGN feedback self-regulation is a compelling mechanism for synchronizing BH and galaxy growth. However, it has also been proposed that BH--galaxy correlations may arise as a non-causal consequence of hierarchical merging \citep{Peng2007,Hirschmann2010,JahnkeMaccio2011} or the result of a common gas supply for BH growth and star formation regulated by gravitational torques \citep{Escala2007,Angles-Alcazar2013,Angles-Alcazar2015,Cen2015,Angles-Alcazar2017a,Soliman2023}. It is likely that both fuelling and feedback play a key role in the co-evolution of BHs and galaxies, but simulation predictions are often affected by factors such as limited resolution and uncertainties in subgrid physics \citep{Somerville&Dave2015}. Detailed comparisons to observations are thus crucial to constrain current AGN self-regulation models. 

Despite their long-term coherence, BH--galaxy scaling relations cannot directly reveal the physical processes that mediate BH fuelling and feedback on short timescales. Addressing this requires high-resolution observations of the gas dynamics in the central tens to hundreds of parsecs, where angular momentum transport, accretion, and outflow launching occur. Optical and near-infrared integral-field unit (IFU) spectroscopy of nearby AGN on 8--10\,m class telescopes (e.g. Gemini/GMOS and NIFS, VLT/MUSE and SINFONI, Keck/OSIRIS), as well as NIRSpec and MIRI-MRS on JWST, now routinely achieves spatial resolutions of tens of parsecs, and has revealed a rich phenomenology of both feeding and feedback on nuclear scales (see \citealt{Harrison2024} for review). These campaigns find inflows of ionized, atomic, and molecular gas along bars, nuclear spirals, and circumnuclear discs on $\sim$10–100\,pc scales that can fuel the SMBH, and biconical ionized outflows often aligned with ionization cones and, in some cases, radio jets \citep[e.g.,][]{Storchi-Bergmann2010,Riffel2013,Riffel2015,Venturi2018,Shimizu2019,Audibert2023,Dasyra2024,Davies2024,HermosaMunoz2024,Zhang2024}. Together, such studies demonstrate that inflow and outflow signatures frequently coexist in the inner $\sim$10–10$^3$\,pc of active galaxies \citep{Diniz2015,Lena2015}. However, resolving the cold molecular gas, the primary fuel for both star formation and SMBH accretion, requires millimeter interferometry. Recent Atacama Large Millimeter Array (ALMA) surveys of nearby Seyferts at $\sim$10\,pc resolution have played a central role uncovering turbulent circumnuclear discs, molecular tori, and multiphase outflows that provide crucial insight into how AGN interact with their immediate environment \citep[e.g.,][]{Garcia-Burillo2014,Garcia-Burillo2016,Alonso-Herrero2018,Alonso-Herrero2019,Combes2019,Combes2026}.

The Galaxy Activity, Torus, and Outflow Survey \citep[GATOS;\footnote{\url{https://gatos.myportfolio.com}}][]{Alonso-Herrero2021,Garcia-Burillo2021} has played a central role in this area by mapping the distribution and kinematics of molecular gas within the circumnuclear discs of nearby Seyfert galaxies. \citet{Garcia-Burillo2021} present ALMA observations of the CO(3–2) and HCO$^{+}$(4–3) lines resolving dusty molecular tori with diameters of $\sim$40\,pc and gas masses of $\sim$$10^{6}\,\Msun$ in 19 AGN, including six targets from the Nuclei of Galaxies (NUGA) survey \citep{Audibert2019,Combes2019}. This work introduced a concentration index ${\rm CCI}\equiv{\rm log}_{10}(\Sigma_{\rm 50\,pc}/\,\Sigma_{\rm 200\,pc})$ based on the molecular gas surface densities within radii of 50\,pc and 200\,pc ($\Sigma_{\rm 50\,pc}$ and $\Sigma_{\rm 200\,pc}$, respectively) as a diagnostic of the compactness of the nuclear gas reservoir. They found tentative evidence that the molecular gas concentration decreases with increasing 2-10\,keV X-ray luminosity ($L_{\rm X}$; hereafter), possibly indicating the impact of AGN feedback dispersing or heating the nuclear gas.
This trend was significantly reinforced in \citet{Garcia-Burillo2024}, which extended the sample to 64 nearby disc galaxies (45 AGN and 19 non-AGN), reporting a bimodal relationship between CCI and $L_{\rm X}$ with two distinct branches: (1) the AGN ``build-up'' branch (with $L_{\rm X}\lesssim 10^{41.5}\,{\rm erg}\,{\rm s}^{-1}$) characterised by high gas concentrations and steeply peaked radial profiles, suggesting efficient fuelling, and (2) the AGN ``feedback branch'' (with $L_{\rm X} \gtrsim 10^{41.5}\,{\rm erg}\,{\rm s}^{-1}$) showing markedly lower gas concentrations and flatter or even inverted gas profiles, interpreted as evidence for depletion or redistribution of nuclear gas due to AGN feedback. These trends were observed in both the cold (CO) and hot molecular hydrogen phases, with the hot-to-cold gas mass ratio increasing in higher-luminosity AGN. Such results offer compelling support for the existence of self-regulation cycles operating on nuclear ($r \lesssim 200\,{\rm pc}$) scales in the local Universe, providing strong constraints for BH accretion and feedback models.

However, these observations inherently capture single-epoch snapshots of dynamic processes that likely evolve over timescales much longer than the typical variability of AGN light curves. This poses a challenge for assessing causal relationships or tracking feedback cycles in individual galaxies. Moreover, although the observed trends are statistically significant, they remain sensitive to orientation effects, projection uncertainties, and differences in sample selection and host-galaxy properties. As such, a crucial question remains: \textit{are the concentration–luminosity trends seen in nearby Seyferts a transient coincidence, or do they reflect a fundamental outcome of AGN self-regulation processes operating in the broader context of galaxy evolution?}

To address this question, we turn to high-resolution cosmological hydrodynamic simulations of Milky Way-mass galaxies from the Feedback In Realistic Environments (FIRE)\footnote{\url{http://fire.northwestern.edu}} project \citep{Hopkins2014,Hopkins2018}, specifically the FIRE-3 version of the model including novel prescriptions for BH accretion and AGN feedback \citep{Hopkins2023_fire3}. These simulations are ideally suited for investigating the relationship between BH growth and host-galaxy evolution by providing a complete evolutionary history while resolving the nuclear gas distribution ($<$100\,pc) in a full cosmological setting, including detailed ISM and stellar feedback physics and a subgrid accretion disc reservoir modulating the impact of multi-channel (radiative and mechanical) AGN feedback. The high resolution and detailed physics implemented in FIRE-3 enable us to follow the history of BH fuelling and feedback cycles, trace the response of the circumnuclear gas, and directly compare simulations to observational constraints.

\citet{Byrne2024} analysed a suite of FIRE-3 simulations to investigate the long-term impact of AGN feedback and star formation quenching in massive galaxies, building on a previous AGN feedback parameter survey using FIRE-2 simulations by \citet{Wellons2023}.
In this study, we focus in more detail on the late time evolution of simulated galaxies and BH self-regulation on nuclear scales.  In particular, we aim to reproduce and interpret the empirical trends reported by \citet{Garcia-Burillo2021,Garcia-Burillo2024} within a cosmological galaxy formation framework. We compute gas concentration indices from the simulations in a manner analogous to the observational definition (i.e., using surface densities within 50\,pc and 200\,pc) and correlate them with simulated X-ray luminosities derived from BH accretion rates. 
By comparing simulated galaxies with the observed GATOS sample, we seek to identify whether the concentration–luminosity trends are indeed manifestations of an underlying self-regulation mechanism. If such a correlation emerges in the simulations without being explicitly imposed, it would lend strong support to the interpretation that AGN feedback actively reshapes the nuclear gas distribution. On the other hand, if the correlation does not emerge in the simulations, or if alternative drivers dominate, this would either call for a re-interpretation of the observations or point to shortcomings in the subgrid models and resolution limits that prevent the simulations from capturing the observed behaviour.

The outline of this paper is as follows: \S\ref{sec:methods} provides a brief summary of the galaxy formation framework and our methodology; \S\ref{sec:overview} presents an overview of the simulations; \S\ref{sec:gas_properties} explores the recurrent cycles in AGN fuelling and feedback; \S\ref{sec:correlation} investigates the connection between gas concentration and X-ray luminosity; \S\ref{sec:density_profiles} examines nuclear gas surface density profiles across bins in $L_{\rm X}$; \S\ref{sec:discussion} discusses our results in the context of previous work; and \S\ref{sec:conclusion} provides a summary of our findings and the main conclusions of this work.

\section{Methods} \label{sec:methods}

\subsection{FIRE-3 galaxy formation model}
We analyse a suite of cosmological zoom-in simulations using the ``FIRE-3'' galaxy formation physics model ran with the GIZMO\footnote{\url{http://www.tapir.caltech.edu/~phopkins/Site/GIZMO.html}} hydrodynamics code in meshless finite mass mode \citep{Hopkins2015gizmo}. The FIRE-3 methodology and physical ingredients are detailed in \citet{Hopkins2023_fire3}. These simulations incorporate all standard FIRE-3 physics, including a full range of stellar feedback mechanisms (Type I and II supernovae, OB and asymptotic giant branch stellar winds, photoionization, and radiation pressure) as well as star formation in dense, self-gravitating gas. The high resolution allows us to resolve the multiphase ISM, with Milky Way-mass galaxies showing a broad range of ISM densities ($n_{\rm H}\sim10^{-3}$--$10^{3}\,{\rm cm}^{-3}$) and temperatures ($T\sim 100$--$10^{6}\,{\rm K}$), where the molecular fraction of high density gas ($n_{\rm H}>100\,{\rm cm}^{-3}$) reaches close to one. All runs also include magnetic fields, following the magnetohydrodynamic implementations of \citet{Hopkins&Raives2016} and \citet{Hopkins2016}. In addition, FIRE-3 introduces new optional physics modules compared to previous versions (\citealt{Hopkins2014, Hopkins2018}), notably a subgrid BH model encompassing BH seeding, dynamics, accretion, and feedback, as well as the inclusion of cosmic ray (CR) physics. 

\begin{table}
\hspace{-0.25cm}
\begin{tabular}{c|| |c|c|c|c|} 
Halo        & $M_{\rm halo}$ [$\Msun$]         & $M_{\star}$ [$\Msun$]   & $M_{\rm BH}$ [$\Msun$]  & SFR [$\Msun\,{\rm yr}^{-1}$] \\ \hline\hline
\simi       & $8.6 \times 10^{11}$        & $ 1.4 \times 10^{10}$    & $ 6.01 \times 10^{6}$    & $ 1 $       \\ \hline
\simf       &  $1.1 \times 10^{12}$       & $ 1.5 \times 10^{10}$    & $ 2.79 \times 10^{7}$    & $ 0.01 $       \\ \hline
\simb       &  $1.1 \times 10^{12}$       &  $2.1 \times 10^{10}$    & $ 1.44 \times 10^{7}$    & $ 1.51 $       \\ \hline
\simm       &  $1.3 \times 10^{12}$       &  $ 4.5 \times 10^{10}$    & $ 3.47 \times 10^{6}$    & $ 17 $      \\ \hline
\end{tabular}
\caption{Parameters of the FIRE-3 simulation suite analysed in this work, listing the four Milky Way-mass haloes (`{\bf \texttt{m12}}'s). Columns are: halo mass, stellar mass, black hole mass, and SFR at the final redshift, $z=0$, within 10\% of the virial radius ($0.1\,R_{\rm vir}$).}
\label{tab:sims}
\end{table}

We focus on four Milky Way-mass haloes (Table~\ref{tab:sims}), corresponding to the FIRE-3 BH (no-CR) {\bf \texttt{m12}} runs analysed in \citet{Byrne2024}. \citet{Byrne2024} considered seven {\bf \texttt{m12}} initial conditions in total, but here we restrict ourselves to a homogeneous subset of four runs evolved down to $z=0$ with the same BH/no-CR physics, the same mass resolution (with baryonic and dark matter particle mass resolution of $m_{\rm b} = 7\times 10^{3}\,\Msun$ and $m_{\rm DM} \approx 3.5\times 10^{4}\,\Msun$, respectively), and the same BH seeding prescription. We make this choice so that the differences we analyse are driven by the galaxies themselves, not by differences in the simulation configuration. We restrict the analysis to the no-CR runs rather than the corresponding BH+CR runs because the latter produce a qualitatively different galaxy population at Milky Way masses. In particular, \citet{Byrne2024} found that the current BH+CR model overquenches Milky Way-mass galaxies and tends to produce nearly spherical, early-type systems rather than the disc-like systems most relevant for comparison to nearby Seyfert galaxies. This choice does not imply that CR physics is unimportant: CRs could affect the amount, phase structure, and spatial distribution of gas in the central few hundred parsecs, and therefore may affect the nuclear gas surface densities and concentration measurements analysed below. Our results should therefore be interpreted specifically within the FIRE-3 BH/no-CR galaxy-formation model.

The four simulated galaxies analysed here also lie in the broad stellar- and BH-mass regime of nearby Seyfert galaxies used in the observational comparisons below. As summarised in Table~\ref{tab:sims}, the simulations span $\log_{10}(M_\star/{\rm M_\odot})=10.15$--$10.65$ and $\log_{10}(M_{\rm BH}/{\rm M_\odot})=6.54$--$7.45$. These values place the simulated systems in the mass range of local AGN hosts commonly used for empirical $M_{\rm BH}$--$M_\star$ comparisons \citep[e.g.][]{Reines2015}, and broadly overlap the nearby Seyfert samples analysed by \citet{Garcia-Burillo2021,Garcia-Burillo2024}. We use adaptive gravitational softenings for gas, with a minimum value of $\epsilon_{\rm gas}^{\rm min}\approx 0.14\,{\rm pc}$, and fixed physical softenings for stars and dark matter of $\epsilon_{\star}\approx 5\,{\rm pc}$ and $\epsilon_{\rm DM}\approx 50\,{\rm pc}$, respectively. For dense gas with $n_{\rm H} > 100\,{\rm cm}^{-3}$ the Plummer equivalent softening is $<$10\,pc, which represents the typical resolution for gas that dominates the inflow rate down to the BH.
These simulations include the BH physics module (described below) but no CR physics. We assume a $\Lambda$CDM cosmology with parameters $H_{0}\approx 70\,{\rm km}\,{\rm s}^{-1}\,{\rm Mpc}^{-1}$, $\Omega_{\rm M}\approx 0.3$, and $\Omega_{\rm b}\approx 0.05$, consistent with recent measurements \citep{PlanckCollaboration2020}.

\subsection{Black hole physics}\label{subsec:bhphysics}
In our simulations, BH seeding, dynamics, accretion, and feedback follow the methods presented in \citet{Hopkins2023_fire3}. BHs originate probabilistically from star-forming gas cells that meet strict conditions: extremely high gas surface densities ($\Sigma_{\rm gas}\geq 5000\,\Msun\,{\rm pc}^{-2}$) and low metallicities ($Z_{\rm gas}\leq 0.001\,Z_{\odot}$). Under these conditions, a seed BH particle forms with mass $M_{\rm BH}\approx100\,\Msun$. The gravitational torque-driven accretion model employed is only weakly dependent on $M_{\rm BH}$, where the final BH masses are dominated by accretion and are insensitive to the initial seed mass \citep{Angles-Alcazar2013,Angles-Alcazar2017a,Angles-Alcazar2017c}.
Because the orbital decay of low-mass BHs is not fully resolved, unresolved dynamical friction and unresolved anchoring processes can lead to artificial BH wandering away from the galactic centre \citep[e.g.,][]{Tremmel2015,Chen2022}. We therefore drift low-mass BHs toward the local gravitational-potential minimum at a fraction of the local velocity dispersion, following \citet{Wellons2023}. This numerical treatment is intended to mitigate unresolved orbital decay, while physically low-mass BH seeds may still sink inefficiently in clumpy galaxies \citep{Bellovary2021,Ma2021}.

Each BH with mass $\Mbh$ is surrounded by a subgrid accretion disc of mass $\Mdisk$. Gas from the accretion kernel ($R_{256}$; defined to include the 256 nearest gas resolution elements with upper limit $R_{256}\lesssim 7$\,kpc) accretes onto the disc at a rate $\dot{M}_{\rm AD} = \eta_{\rm acc}\, M_{\rm gas} \,\Omega$, where the efficiency parameter $\eta_{\rm acc}$ is calibrated from high-resolution simulations to represent the effects of gravitational torques driving gas towards the BH due to angular momentum loss to the stellar component \citep{Hopkins&Quataert2011,Hopkins2016,Angles-Alcazar2017a,Angles-Alcazar2017c,Angles-Alcazar2021}. The gas mass $M_{\rm gas}$ and dynamical frequency $\Omega = \sqrt{GM_{\rm tot}/R^3}$ are evaluated within the accretion kernel, where $M_{\rm tot}$ is the total mass within $R_{256}$.
The rate at which the BH accretes from the accretion disc, $\MdotBH$, is determined following a \cite{Shakura1973}-like $\alpha$-disc subgrid prescription:
\begin{equation}\label{eq:disk}
\MdotBH \equiv \frac{M_{\rm AD}}{t_{\rm acc}},
\end{equation}
where, in our simulations,
\begin{equation}\label{eq:tacc}
t_{\rm acc} \equiv 42~{\rm{Myr}} \left[1 + \frac{M_{\rm BH}}{M_{\rm AD}} \right]^{0.4}.
\end{equation}

Here $t_{\rm acc}$ is the adopted unresolved accretion-disc depletion timescale in the FIRE-3 BH model, motivated by a simple analytic $\alpha$-disc prescription rather than chosen specifically to match either the observations or the snapshot cadence \citep{Hopkins2023_fire3,Wellons2023}. The normalization, $42\,{\rm Myr}$, is of the same order as the Salpeter/e-folding time for Eddington-limited BH growth for $\epsilon_{\rm rad}\simeq 0.1$, and sets the characteristic minimum depletion time of a massive unresolved disc reservoir. Thus, when $M_{\rm AD}\sim M_{\rm BH}$, the model allows the BH to approach the Eddington accretion regime, while for smaller disc reservoirs the bracketed factor increases $t_{\rm acc}$ and suppresses $\MdotBH$ \citep{Hopkins2023_fire3}. Modest changes to the normalization or slope have only small effects on the model behaviour, while the key feature is that the finite disc-depletion time smooths rapid fluctuations in BH feeding and allows AGN feedback to persist even as nuclear gas is being evacuated, which helps produce luminous AGN phases with depleted central gas cavities similar to those inferred in nearby systems \citep{Garcia-Burillo2024}.

Accreting BHs inject energy and momentum into their surrounding ISM through both radiative and mechanical feedback.  
Radiation transport is modeled using the same LEBRON approximation employed for stellar radiative feedback in the FIRE simulations \citep{Hopkins2020}. The accretion disc emits radiation with a bolometric luminosity $\dot{E}_{\rm rad} \equiv L_{\rm bol} = \epsilon_{\rm rad}\, \dot{M}_{\rm BH}\, c^2$,
where $c$ is the speed of light and the radiative efficiency is fixed to $\epsilon_{\rm rad} = 0.1$. 
In the LEBRON model, local photoionization approximately follows a Str\"omgren-sphere construction around the BH, and the absorbed short-range single-scattering radiation-pressure term is estimated from the local BH-kernel optical depth. The unabsorbed and re-radiated luminosity is propagated to larger scales in the optically-thin approximation and contributes photoionization, Compton heating, and radiation pressure \citep{Hopkins2020,Hopkins2023_fire3}.

Mechanical (or kinetic) feedback is implemented via the injection of new high-resolution gas elements (or wind particles) in the immediate vicinity of the BH \citep[see][]{Richings2018, Torrey2020, Su2021, Cochrane2023, Mercedes-Feliz2023, Mercedes-Feliz2024}. 
Importantly, mechanical feedback is thus not implemented by depositing energy into pre-existing gas resolution elements within the BH kernel, which can bias the impact of feedback in the direction of the cold ISM component.
The high-resolution, spawned wind particles have mass $m_{\rm wind}\sim 100\,\Msun$, approximately two orders of magnitude smaller than the typical baryonic particle mass in these simulations. 
Their initial launched positions are offset from the BH by a small scale limited by the gravitational softening and the separation to nearby gas. The wind particles are launched symmetrically in bipolar pairs along the angular momentum axis of the unresolved accretion disc, conserving linear momentum, and subsequently propagate and interact hydrodynamically with the surrounding multiphase ISM.
Wind particles are launched with a fixed velocity of $v_{\rm wind} = 3,000\,{\rm km\,s^{-1}}$ and a total mass outflow rate equal to the accretion rate, i.e., $\dot{M}_{\rm wind} = \dot{M}_{\rm BH}$. In this adopted subgrid prescription, the unresolved disc wind is therefore assigned a mass-loading factor of unity relative to the BH accretion rate. 
Once wind particles have transferred most of their energy and momentum to the surrounding gas, they are allowed to merge with nearby regular gas elements while conserving mass, momentum, and energy.
These outflows are initially collimated but, upon interaction with the surrounding ISM, they undergo shock heating and preferentially propagate along paths of least resistance, eventually filling a wide solid angle. 

In the simulations analysed here, the radiative and mechanical AGN feedback prescriptions do not transition between feedback modes as a function of Eddington ratio. The radiative efficiency, wind velocity, and wind mass-loading prescription are fixed as described above.
Importantly, AGN feedback is implemented independently of the definition of the BH kernel. Wind particles spawned in the immediate vicinity of the BH propagate and affect the surrounding gas self-consistently regardless of the gas particle location relative to the BH kernel.  Similarly, radiation is injected at the BH location and transported outward by the LEBRON approximation, with gas particles impacted by the corresponding radiative effects regardless of the definition of the BH kernel. Therefore, $R_{256}$ does not define the set of pre-existing gas elements that directly receive feedback, nor does it prescribe the radius of the resulting cavity.

In addition to snapshots with full simulation data available at $\Delta t \sim 25$--40\,Myr intervals ($\Delta t \sim 1\,{\rm Myr}$ in the last 100 Myr), reflecting the standard full-snapshot output cadence, we use BH-specific outputs to perform high time resolution analyses of the evolution of nuclear gas densities and the resulting BH accretion rates \citep{Angles-Alcazar2017c}.

\begin{figure*}
\includegraphics[width = \textwidth]{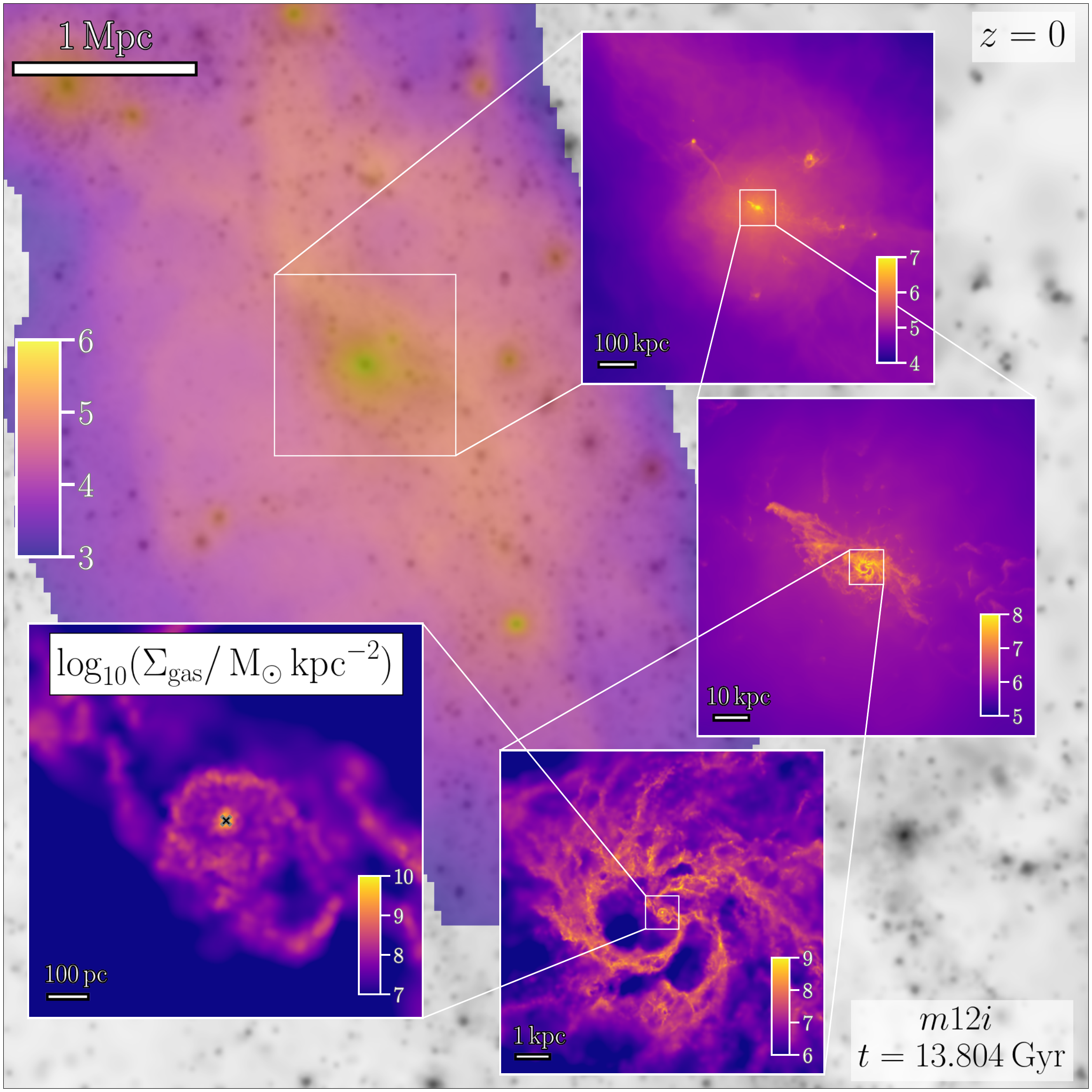}
\vspace*{-5mm}
\caption{Projected gas mass surface density distribution across scales in one of our simulations (\simi) at $z = 0$. The colourbars encode surface density on a logarithmic scale with their range optimized to the different dynamic range in each panel. In the largest panel, we show the large-scale structure surrounding the galaxy on $\sim$10\,Mpc scales (with the dark matter distribution illustrated in gray), zooming-in progressively down to the central 1\,kpc of this Milky Way-mass galaxy. The simulation captures $>$5 orders of magnitude variation in gas surface density ($\Sigma_{\rm gas}\sim 10^{4-9}\,\Msun\,{\rm kpc}^{-2}$) and a dynamic range spanning over five orders of magnitude in spatial scale.}
\label{fig:zoomin_overview_map} 
\end{figure*}

\begin{figure*}
{\includegraphics[width = 0.9\textwidth]{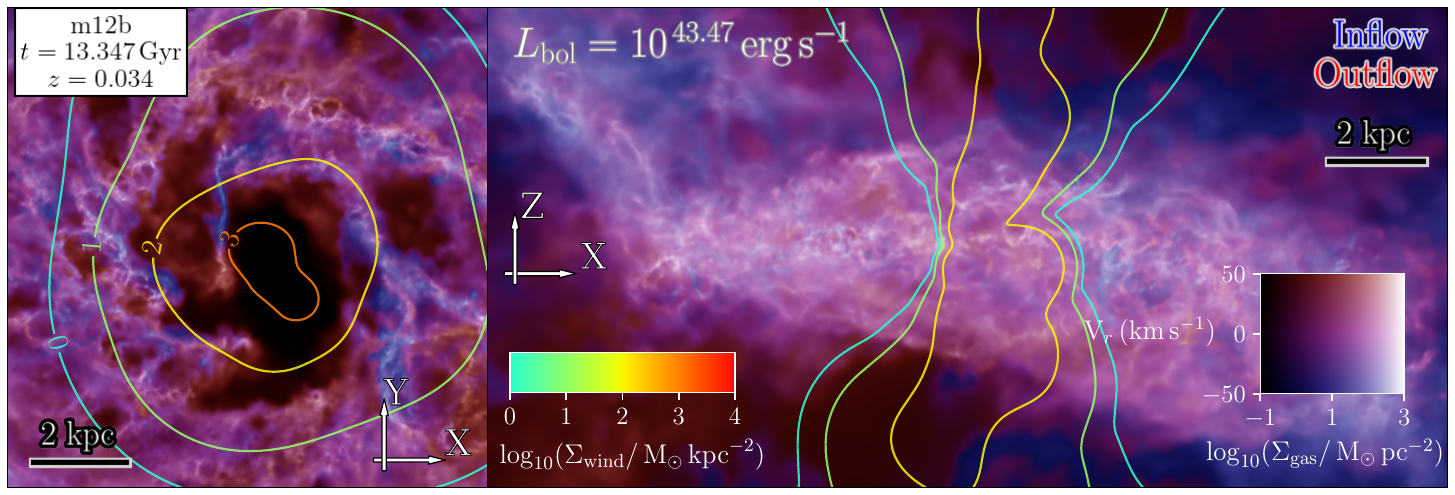}\vspace{-2mm}
\includegraphics[width = 0.9\textwidth]{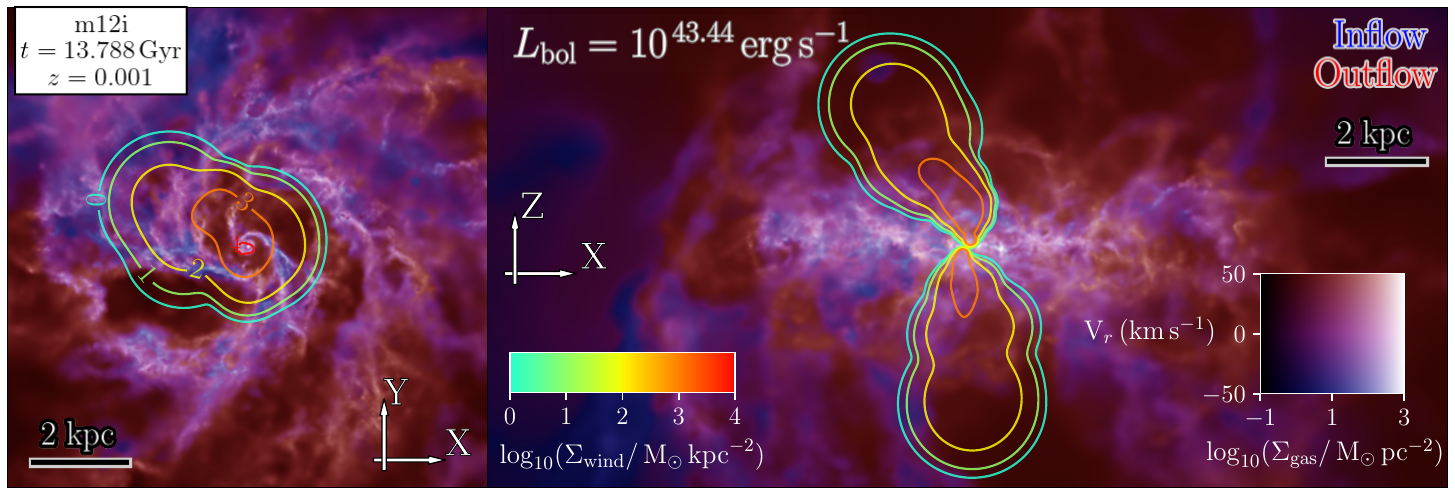}\vspace{-2mm}
\includegraphics[width = 0.9\textwidth]{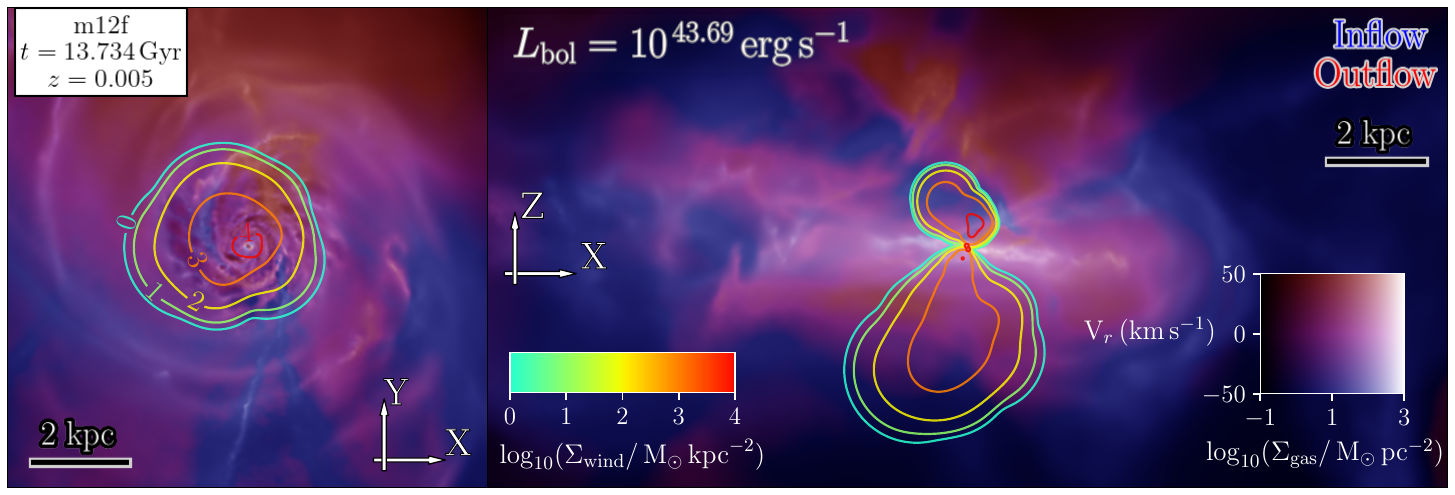}\vspace{-2mm}
\includegraphics[width = 0.9\textwidth]{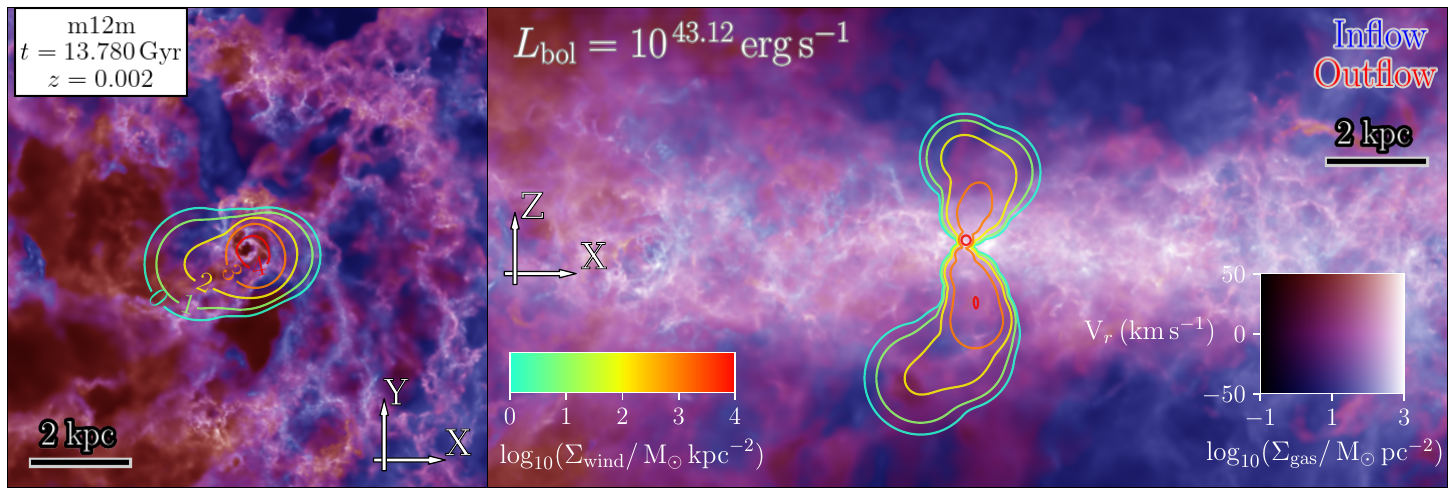}}
\vspace*{-2mm}
\caption{Projected gas mass surface density distribution for the four simulated Milky Way-mass galaxies analysed in this study (from top to bottom: \simb, \simi, \simf, and \simm) at representative times with different levels of AGN activity. The face-on projection (left) shows the central 5\,kpc region, while the edge-on projection (right) shows the central 10\,kpc region. Inflowing gas is represented in blue hues, outflowing gas is represented in red hues, and the colour value and saturation correspond to the gas surface density logarithmically scaled. Contours show the projected mass distribution of wind particles, illustrating the propagation of AGN-driven winds.
All galaxies are disc dominated and show a variety of sizes and morphological features, with AGN winds preferentially escaping along the polar direction after opening a central cavity.
}
\label{fig:all_overview} 
\end{figure*}

\section{Overview of Simulations} \label{sec:overview}

Figure~\ref{fig:zoomin_overview_map} shows the face-on projected gas mass surface density for one of our four simulated galaxies (\simi) at $z=0$, illustrating the dynamic range of our simulations on different scales, from $\sim$10\,Mpc down to the central kpc. 
The background panel shows the full extent of the zoom-in region. The main halo/galaxy ($ M_{\rm halo}=8.6\times 10^{11}\,\Msun$ and $ M_{\star}=1.4\times 10^{10}\,\Msun$; centred within the white box) is surrounded by high-density regions corresponding to satellite galaxies. The remaining gas distribution traces the filamentary structure through which the main galaxy accretes material.

The subsequent two panels, which zoom into the central 1\,Mpc and 100\,kpc regions, reveal the complex gas dynamics arising from the interplay between filamentary accretion from the intergalactic medium (IGM) onto the galactic disc, cooling of hot halo gas, and galactic winds from the central galaxy and its satellites \citep{Angles-Alcazar2017b,Hafen2019,Hafen2020}. The 100\,kpc panel highlights the inhomogeneous, clumpy structure of the circumgalactic medium (CGM), consistent with a multiphase halo containing cool gas substructure on large scales. The final panel, which focuses on the central 1\,kpc of the galaxy, reveals the intricate structure of the nuclear gas disc and the immediate surroundings of the central SMBH. The high gas surface densities in this region ($\Sigma_{\rm gas}\sim 10^{8-10}\,\Msun\,{\rm kpc}^{-2}$) demonstrate that the simulation can resolve the dense gas reservoirs that serve as the direct fuel for both nuclear star formation and BH accretion. 

\begin{figure*}
\centering
\centerline{\includegraphics[width = \columnwidth]{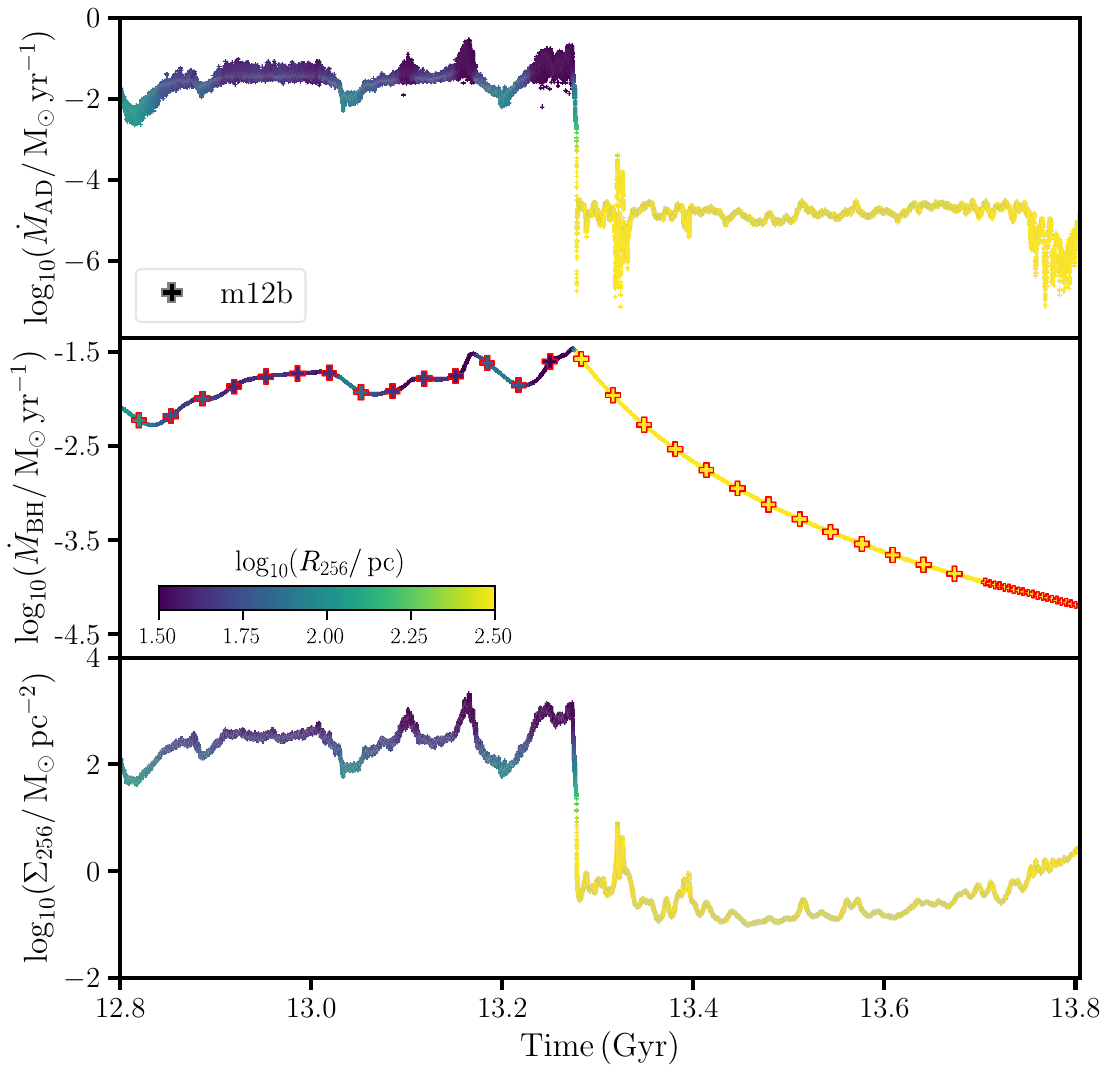}
\includegraphics[width = \columnwidth]{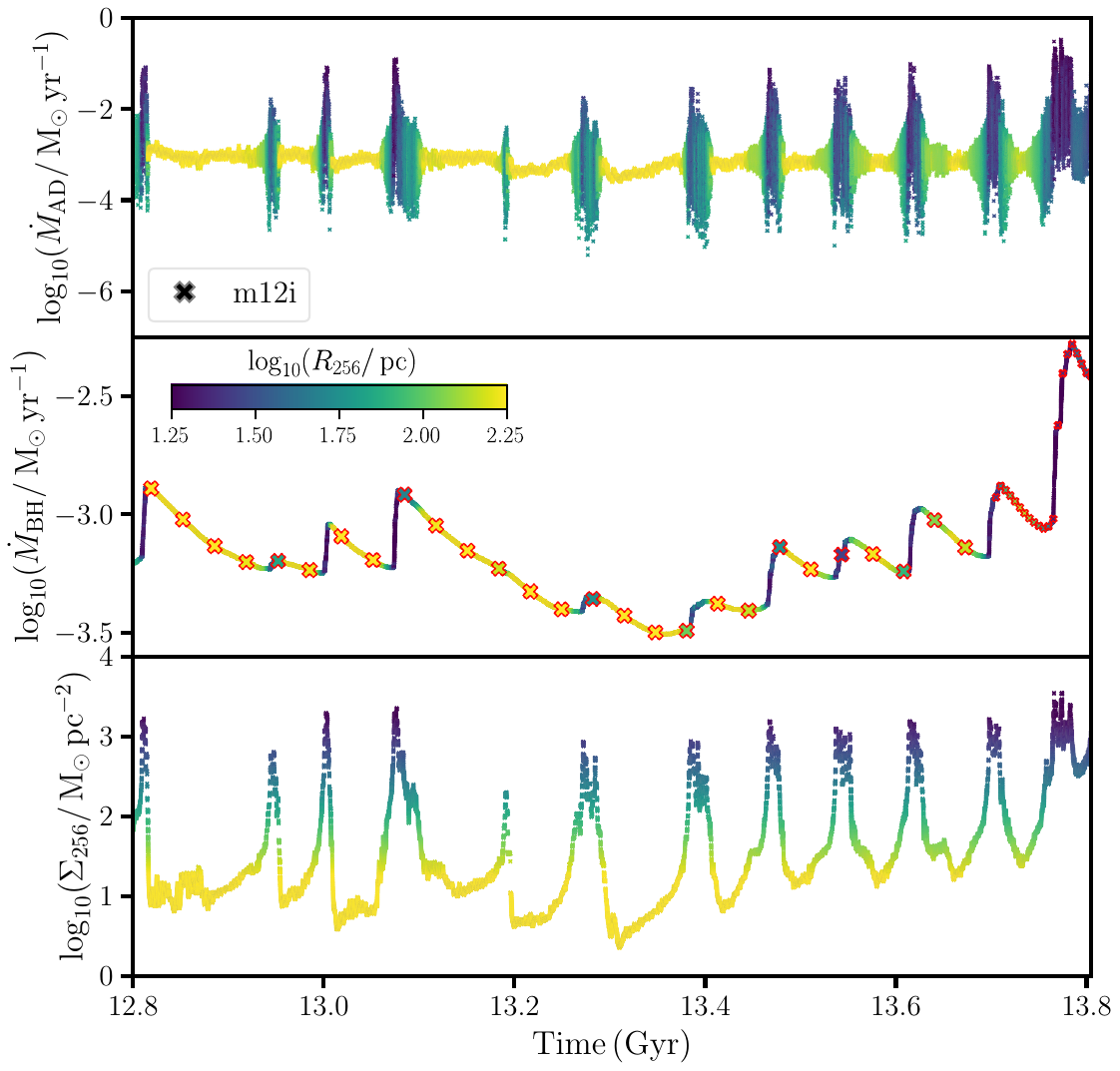}}
\centerline{\includegraphics[width = \columnwidth]{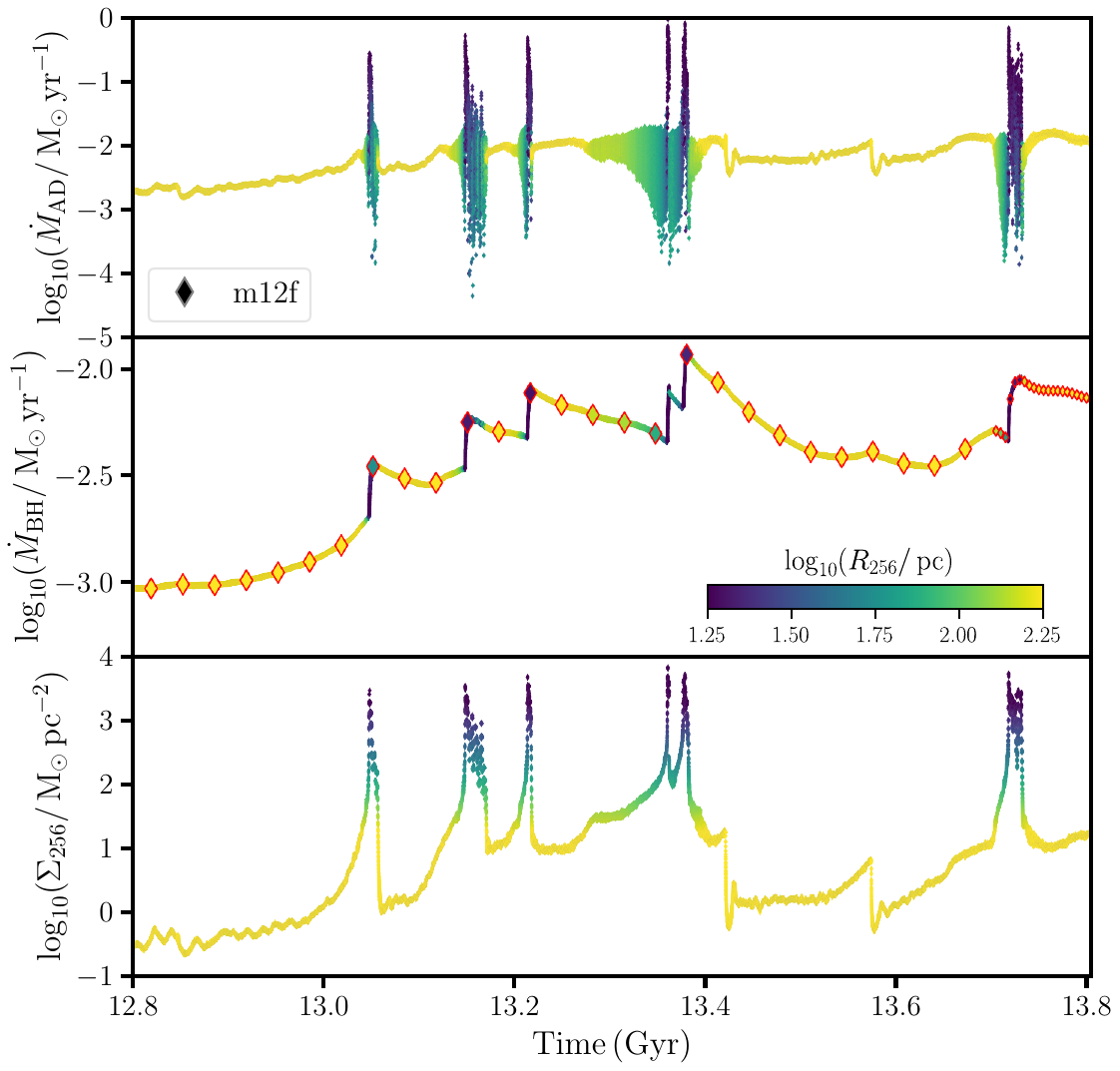}
\includegraphics[width = \columnwidth]{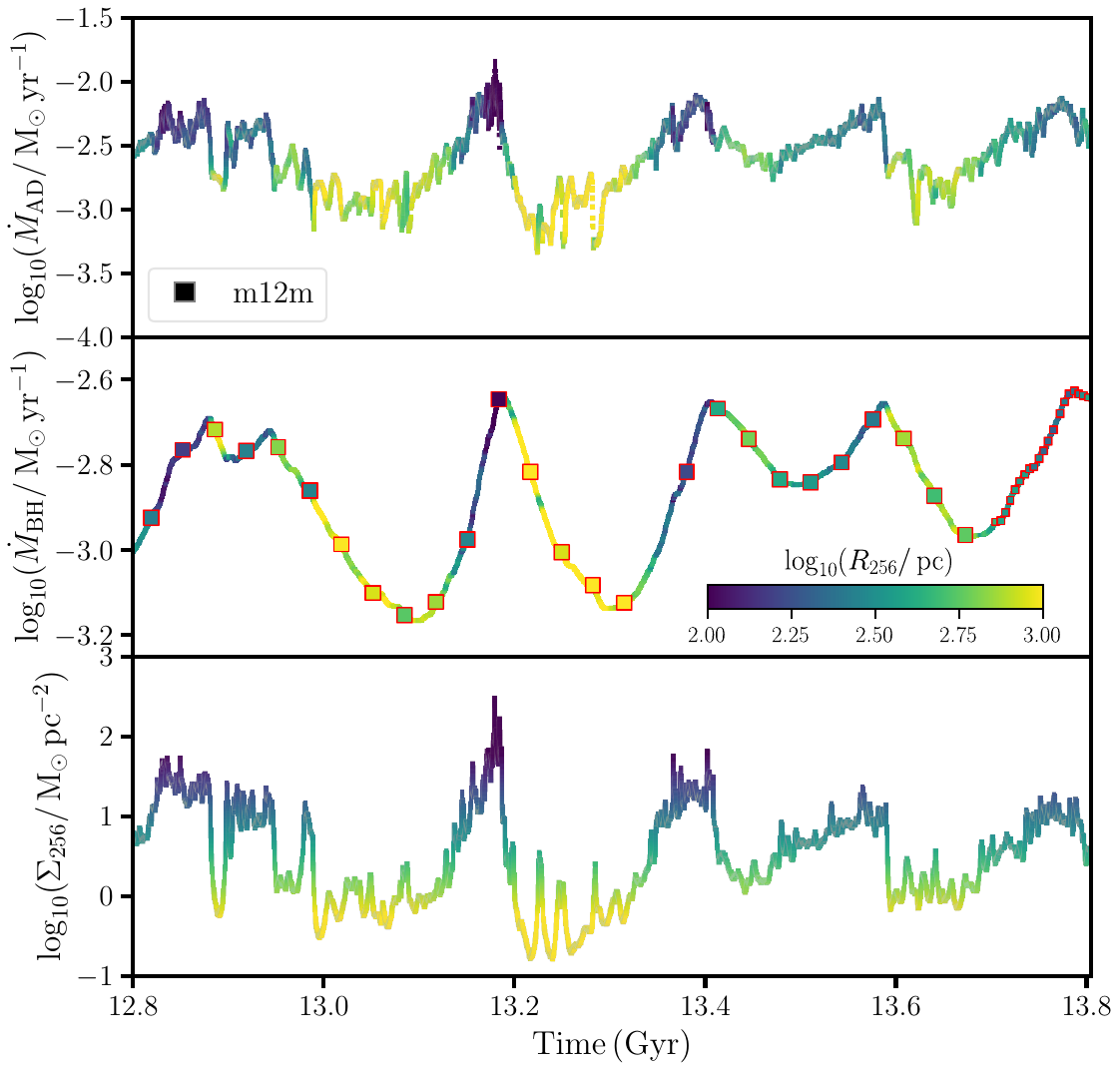}}
\vspace*{-2mm}
\caption{For the four simulations, we show the time evolution of the inflow rate from galaxy scales down to the subgrid accretion disc ($\dot{M}_{\rm AD}$; top panel), the accretion rate onto the BH ($\dot{M}_{\rm BH}$; middle panel), and the gas mass surface density within the BH kernel ($\Sigma_{\rm 256}$; bottom panel). The colour scale for each line indicates the size of the BH kernel ($R_{\rm 256}$), defined as the radius that encompasses the nearest 256 gas particles from the BH. The points outlined in red in the middle panels correspond to the times for which we have full simulation snapshots available, while the remaining data correspond to the high-frequency, BH-specific data outputs. 
We show the last Gyr of evolution for each galaxy, including multiple AGN fuelling and feedback cycles. Peaks in nuclear gas surface density $\Sigma_{\rm 256}$ correlate with a rapid increase of the inflow rate onto the disc $\dot{M}_{\rm AD}$, driving higher BH accretion rate $\dot{M}_{\rm BH}$ and correspondingly stronger AGN winds. The subgrid accretion disc reservoir can sustain elevated $\dot{M}_{\rm BH}$ rates while AGN winds evacuate gas from nuclear region, decreasing $\Sigma_{\rm 256}$ and suppressing $\dot{M}_{\rm AD}$ until the next fuelling event, in repeated AGN fuelling and feedback episodes.  
}
\label{fig:all_timeplots} 
\end{figure*}

Figure~\ref{fig:all_overview} shows the face-on (left) and edge-on (right) projected gas mass surface densities for our four simulated galaxies at representative times with different levels of AGN feedback.
We focus on the central 5\,kpc (10\,kpc wide) region for the face-on maps, while for the edge-on projections we expand the view to a 20\,kpc wide rectangular region. 
The images are colour-coded by the mass-weighted radial velocity to distinguish between inflowing and outflowing gas components, highlighting the diverse morphologies, dynamical states, and spatial extents of the galaxies. 
We use a hue-saturation-value (HSV) transformation where:
(i) Hue represents the radial velocity, encoding inflows and outflows across a spectrum from blue (inflow) to red (outflow);
(ii) Saturation is modulated by the gas mass surface density to enhance visibility of denser regions, but bounded away from zero so that the radial-velocity information is not washed out in the highest-density regions;
(iii) Value is derived from the gas mass surface density to control brightness and contrast in the image.
The radial velocity is defined relative to the centre of mass velocity of stars within the central 1\,kpc of the galaxy. We use the terms ``inflowing'' and ``outflowing'' in a local kinematic sense, corresponding to negative and positive radial velocities relative to the galaxy centre, respectively. These velocities should not be interpreted as an escape criterion: the mass-weighted projected velocities shown here indicate the direction of bulk gas motion in the mapped gas, but do not by themselves imply that the gas is unbound from the galaxy or halo. The contours represent the wind mass surface density and therefore show the projected location and extent of AGN wind material in these galaxies (based on the AGN wind particles identified in each snapshot; see \S\ref{subsec:bhphysics} for details). Because the displayed velocity is mass-weighted along the projection direction, the colour scale is dominated by the denser ISM and should not be interpreted as the velocity of the launched AGN wind material itself. The feedback signatures in these maps are therefore identified from the combination of positive radial motions, wind-particle surface-density contours, central gas cavities, and low-surface-density extraplanar lobes.

All four Milky Way-mass galaxies develop disc-like structures by $z=0$, with differences in their morphology and gas distribution arising from their different halo growth histories and variations in gas supply, recent starbursts, and luminous AGN phases.
Simulated galaxies exhibit a range of structures: they can be compact with  well-defined discs and smaller projected vertical extent in the edge-on views (\simi~and \simf) or more extended with broader gas distributions and larger projected vertical extent (\simb~and \simm).
These representative snapshots show extended biconical outflows of various sizes, extending from $\sim$2\,kpc (\simf~and \simm) to $>$10\,kpc (\simb) above and below the plane of the galaxy. The gas kinematics within these wind-associated lobes are heterogeneous: some lobes are dominated by positive radial velocities, while others contain mixed or negative-radial-velocity gas, consistent with asymmetric wind--ISM coupling, turbulent motions, or gas that has slowed and begun to return. AGN feedback affects galaxies over a range of scales, but its clearest signatures in these maps are the central cavities, the wind contours, and the lower-surface-density extraplanar outflow lobes, rather than a strong radial-velocity signature in the highest-surface-density disc gas.

Central cavity sizes vary widely, with a prominent kpc-scale opening driven by strong AGN winds in \simb~while much smaller cavities ($\lesssim$10--100\,pc) are seen in the other galaxies. Winds propagate further along paths of least resistance, perpendicular to the plane of the galaxy, but the expansion is often asymmetric due to the inhomogeneous, clumpy ISM \citep[see also][]{Gabor2014,Torrey2020,Mercedes-Feliz2023,Sivasankaran2025}. This is especially evident in the edge-on maps: in \simi, the upper lobe is visibly tilted relative to the lower lobe, while in \simf\ the lower lobe extends farther from the disc than the upper lobe. These differences indicate that, although the winds are launched preferentially near the BH rotation axis, their large-scale propagation is shaped by the surrounding gas distribution and need not remain axisymmetric. In all cases, inflowing and outflowing gas can coexist within the central kpc of the galaxies, with the inflowing component generally concentrated near the disc plane and inner disc structures, while the outflowing component is more prominent in the extraplanar, wind-associated lobes.

\section{AGN fuelling and feedback cycles} \label{sec:gas_properties}

Figure~\ref{fig:all_timeplots} shows the time evolution of the gas inflow rate down to the subgrid accretion disc ($\dot{M}_{\rm AD}$; top panel), the actual BH accretion rate ($\dot{M}_{\rm BH}$; middle panel), and the gas mass surface density within the BH kernel ($\Sigma_{\rm 256}$; bottom panel) over a time window of 1\,Gyr for all four simulations. To highlight cycles of BH self-regulation, the colour scale for each simulation corresponds to the kernel radius $R_{256}$ within which $\Sigma_{\rm 256}$ and $\dot{M}_{\rm AD}$ are evaluated. We determined $R_{256}$ in post-processing based on the kernel-weighted gas density and the average mass of 256 gas resolution elements (assuming constant density inside of the kernel). Since $R_{256}$ expands or contracts responding to the gas distribution around the BH, it is a good indicator of the impact of AGN winds evacuating gas from the central region. We focus on the last 1\,Gyr of evolution for each simulation to illustrate the physical properties and evolution of Seyfert galaxies in the local Universe.

All simulations show substantial variation of nuclear gas surface density over time ($\Sigma_{\rm 256} \sim 0.1$--$10^{4}\,\Msun$\,pc$^{-2}$), undergoing up to four orders of magnitude change in $\Sigma_{\rm 256}$ on timescales $\lesssim$10\,Myr. Periods of increasing gas nuclear density correspond to decreasing BH kernel radius (as indicated by the colour scale) and clearly correlate with peaks in $\dot{M}_{\rm AD}$, indicating phases of increased gas availability in the nuclear region and rapid gas inflow toward the disc driven by gravitational instabilities. We emphasize that a fuelling episode does not imply monotonic growth of $\dot{M}_{\rm AD}$ at every instant. Because $\dot{M}_{\rm AD}$ is evaluated from the instantaneous gas and stellar properties within the adaptive BH kernel, it can fluctuate by $\sim$1--2 dex even within an overall fuelling-dominated phase. This short-timescale modulation appears more pronounced in \simi\ and \simf\ than in \simm\ or \simb, which show smoother evolution, possibly reflecting differences in how clumpy and intermittent the nuclear gas supply is in each system.

High disc feeding rates are rapidly followed by an increase in BH accretion rate (Equation~\ref{eq:disk}), and these episodes are often followed by a rapid depletion of the nuclear gas reservoir, seen as an expanding $R_{256}$ and decreasing $\Sigma_{\rm 256}$. We interpret this behaviour as part of a feedback-regulated fuelling cycle, but the temporal correlation alone does not uniquely identify the feedback channel responsible for removing the gas. In particular, the same dense nuclear gas that drives large $\dot{M}_{\rm AD}$ and subsequently increases $\dot{M}_{\rm BH}$ can also trigger enhanced central star formation. Stellar feedback from the associated starburst may contribute to, or in some cases dominate, the evacuation of the nuclear reservoir. As the reservoir is depleted, the gas inflow rate down to the disc subsequently decreases. The BH accretion rate begins to decline as well, but the subgrid accretion disc acts as a gas reservoir that can maintain elevated $\dot{M}_{\rm BH}$ over longer timescales, effectively smoothing the BH response to rapid fluctuations in $\dot{M}_{\rm AD}$. This lagged response allows BH accretion to remain elevated after the instantaneous nuclear inflow has already declined, making AGN feedback a plausible contributor to the continued clearing of the central region. However, because the depletion episodes can also coincide with delayed stellar-feedback timescales from recent central star formation, these diagnostics do not by themselves prove that AGN feedback is the sole driver of the cavity formation. 

These fuelling--depletion cycles are seen in all simulated galaxies, with characteristic timescales of order $\sim$10--100\,Myr, but we identify significant differences depending on the properties and physical state of the host galaxy. Since all four simulations use the same physics, numerical resolution, and accretion/feedback prescriptions, these differences should not be interpreted as the result of a varied model parameter. Instead, they reflect differences in the gas-supply histories produced by the initial conditions.
For example, while the central BH in \simi~undergoes $>$10 accretion and feedback cycles (lasting $\sim$10--100\,Myr) during the last Gyr down to $z=0$, galaxy \simb~experiences a stronger and more extended fuelling episode, followed by a rapid depletion of the central gas reservoir and a long-lived reduction in nuclear activity lasting several hundred Myr (see also Figure~\ref{fig:all_overview}; top panel). The larger-scale gas reservoir is not depleted in the same way, suggesting that the event is localized to the central region rather than reflecting galaxy-wide gas removal. Thus, \simb\ appears to be a case where the galaxy assembly history allows gas to remain available to the nucleus for longer before the central reservoir is disrupted.

\begin{figure}
\includegraphics[width = \columnwidth]{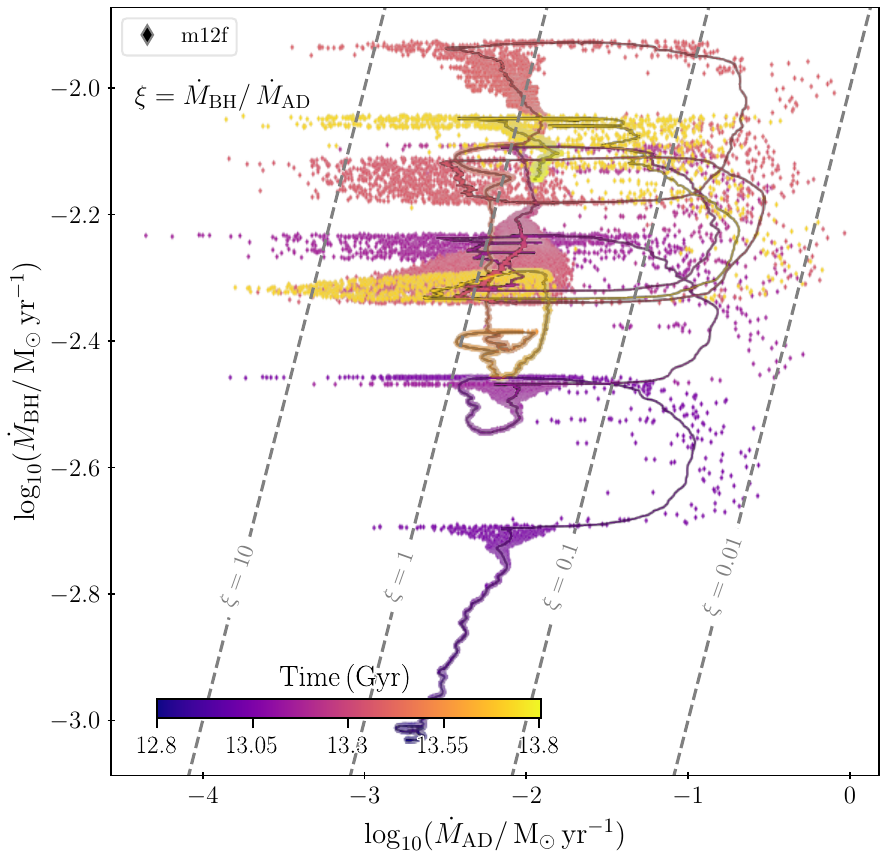}
\vspace*{-5mm}
\caption{The $\dot{M}_{\rm AD}$--$\dot{M}_{\rm BH}$ diagram for simulation \simf, showing the inflow rate onto the accretion disc ($\dot{M}_{\rm AD}$) versus the accretion rate onto the BH ($\dot{M}_{\rm BH}$). Points show individual time-steps throughout the simulation while the solid line is a running average over 1\,Myr. The colour scale denotes time evolution, focusing on the last 1\,Gyr of this galaxy. The dashed lines represent different ratios $\xi \equiv \dot{M}_{\rm BH}/\dot{M}_{\rm AD}$, which measure how much of the disc's inflow ends up in the BH.
AGN fuelling and feedback cycles appear as loops in this diagram: 
higher $\dot{M}_{\rm AD}$ $\rightarrow$ higher $\dot{M}_{\rm BH}$ (stronger feedback) $\rightarrow$ lower $\dot{M}_{\rm AD}$ $\rightarrow$ lower $\dot{M}_{\rm BH}$. 
}
\label{fig:m12f_MandM} 
\end{figure}

\begin{figure*}
\centerline{\includegraphics[width=0.325\textwidth]{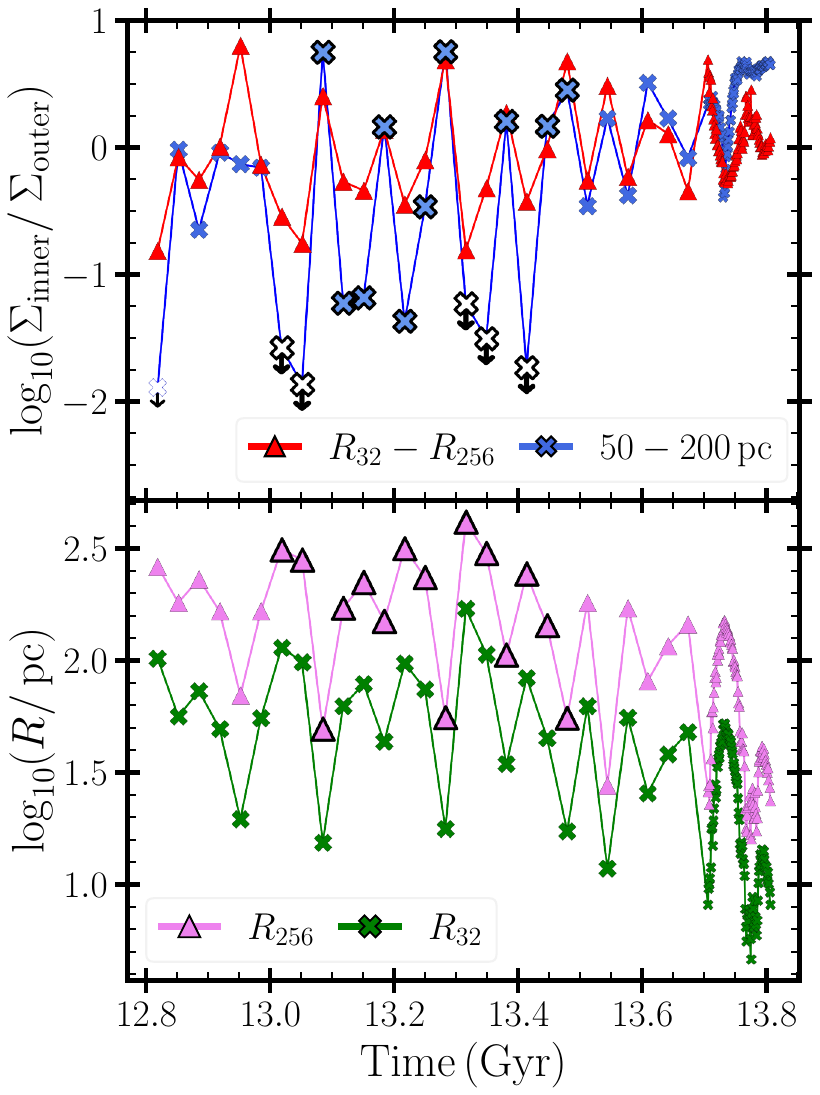}
\raisebox{0.35cm}{\includegraphics[width=0.675\textwidth]{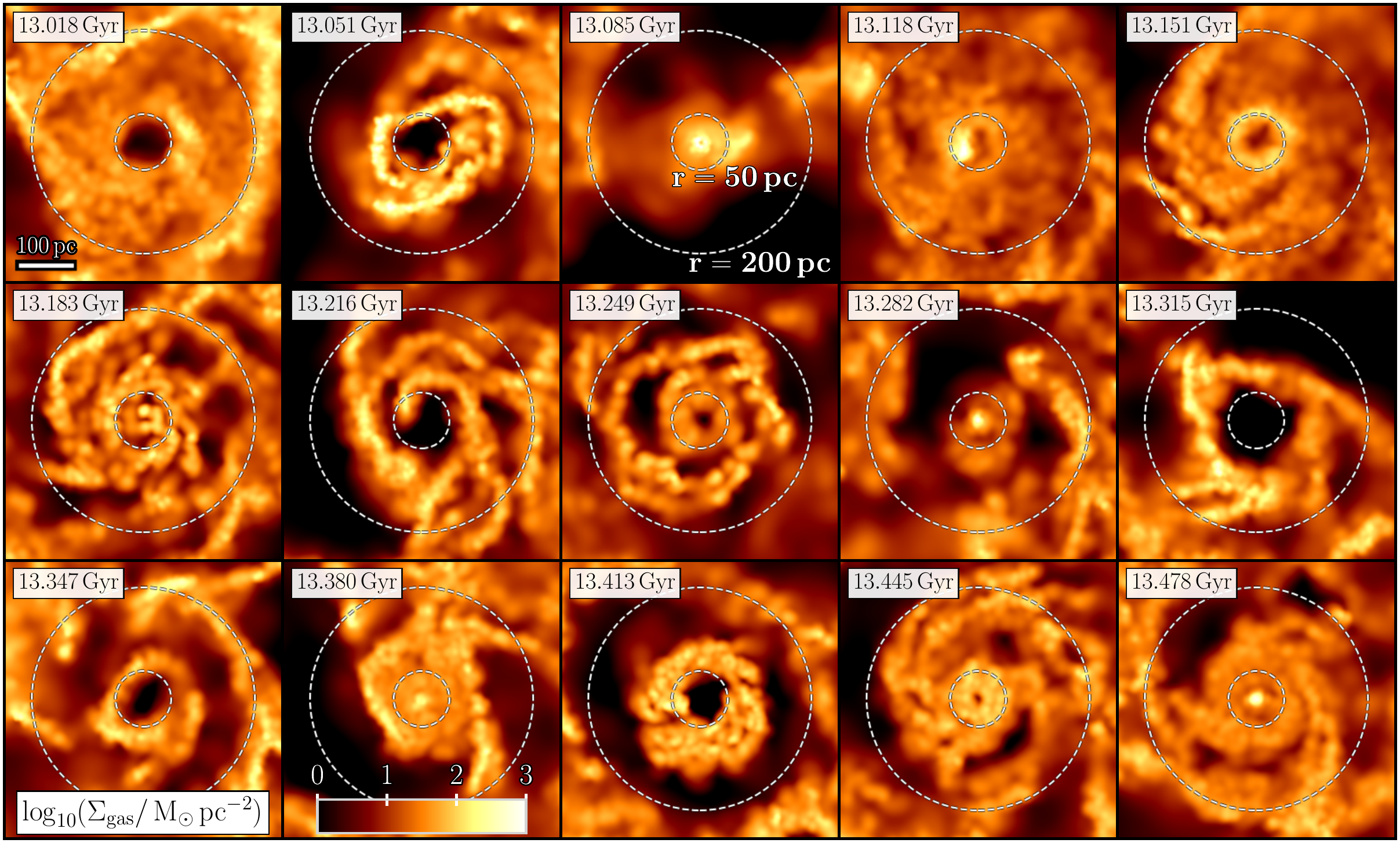}}}
\vspace*{-3mm}
\caption{Left: Time evolution of gas concentration (top) and enclosed radii (bottom) during the last 1\,Gyr of simulation \simi. 
The gas concentration is computed as the ratio of the surface density in the central 50\,pc to 200\,pc region (blue) or between $R_{32}$ and $R_{256}$ (red). Open markers indicate upper limits. Enclosed radii $R_{256}$ (pink) and $R_{32}$ (green) are defined to contain a fixed number of gas resolution elements (either 256 or 32), roughly corresponding to the same amount of gas over time. 
Right: Face-on projected gas mass surface density maps of the central 500\,pc region for representative times indicated as markers highlighted in black in the top left panel. The white dashed circles indicate the central 50\,pc and 200\,pc regions. 
Gas concentration and enclosed radii serve as proxies for the presence or absence of a central cavity, with the simulated galaxy showing a variety of morphological features in the nuclear region.
} 
\label{fig:m12i_multipanel} 
\end{figure*}

\begin{figure*}
\includegraphics[width = 0.9\textwidth]{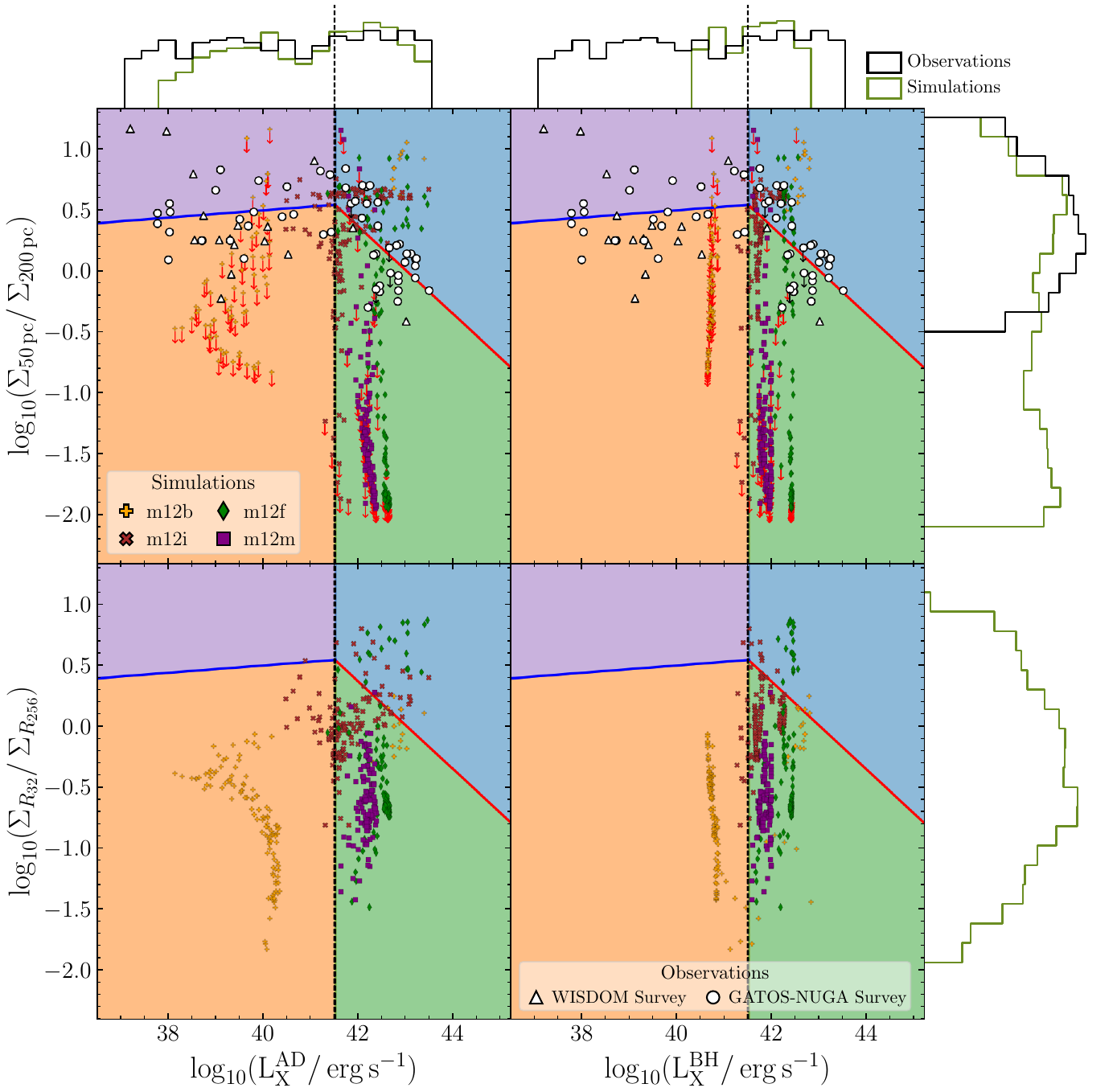}
\caption{Gas concentration as a function of X-ray luminosity over the last Gyr for our four simulated galaxies. Gas concentration is defined as the ratio between the gas surface density within 50\,pc and 200\,pc (top panels) or within $R_{32}$ and $R_{256}$ (bottom panels). The X-ray luminosity is estimated based on the inflow rate down to the accretion disc ($L_{\rm X}^{\rm AD}$; left panels) or based on the accretion rate onto the BH ($L_{\rm X}^{\rm BH}$; right panels).
The white triangle markers correspond to molecular gas concentration and $L_{\rm X}$ measurements for active galaxies in the WISDOM survey \citep{Elford2024}.
The white circle markers correspond to Seyfert galaxies in the GATOS-NUGA survey \citep{Garcia-Burillo2024}. The arrows correspond to upper limits set by the simulations (red) and from observations (black). The two-branch fit for observed Seyfert galaxies as a function of $L_{\rm X}$ is shown as the blue line (AGN {\it build-up phase}) and the red line (AGN {\it feedback phase}), with the black dashed vertical line indicating $L_{\rm X}$ at the inferred transition luminosity. The marginal histograms compare the one-dimensional distributions of the simulated and observed points projected onto each axis: the top histograms show the $L_{\rm X}$ distributions for each luminosity definition, while the right histograms show the gas-concentration distributions for each aperture definition.
Simulated galaxies cover a broadly similar range of parameter space as observations but do not show a clear correlation between gas concentration and $L_{\rm X}$.  
}
\label{fig:comparison_to_obs} 
\end{figure*}

\begin{figure*}
\includegraphics[width = \columnwidth]{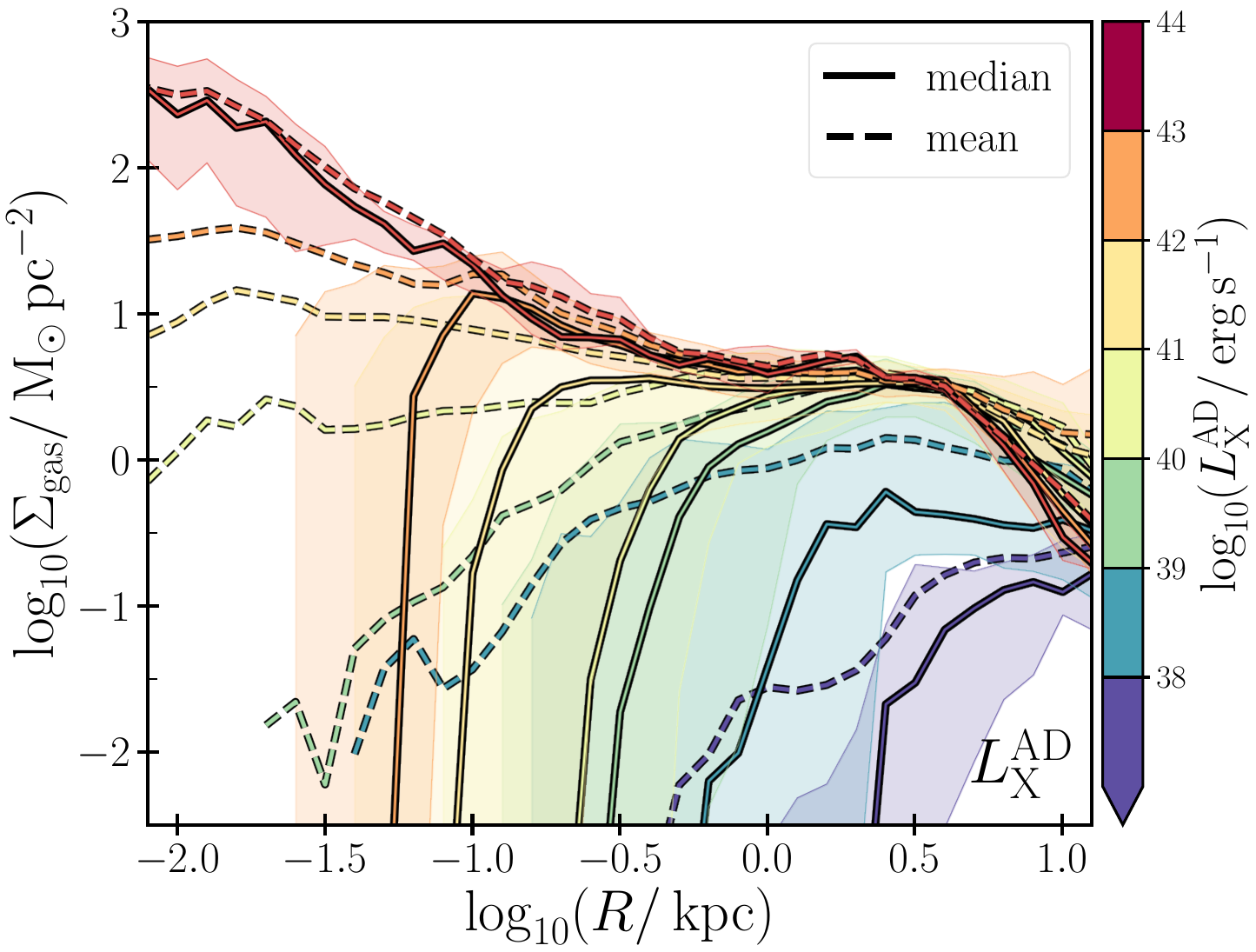}
\hspace{2.5mm}
\includegraphics[width = \columnwidth]{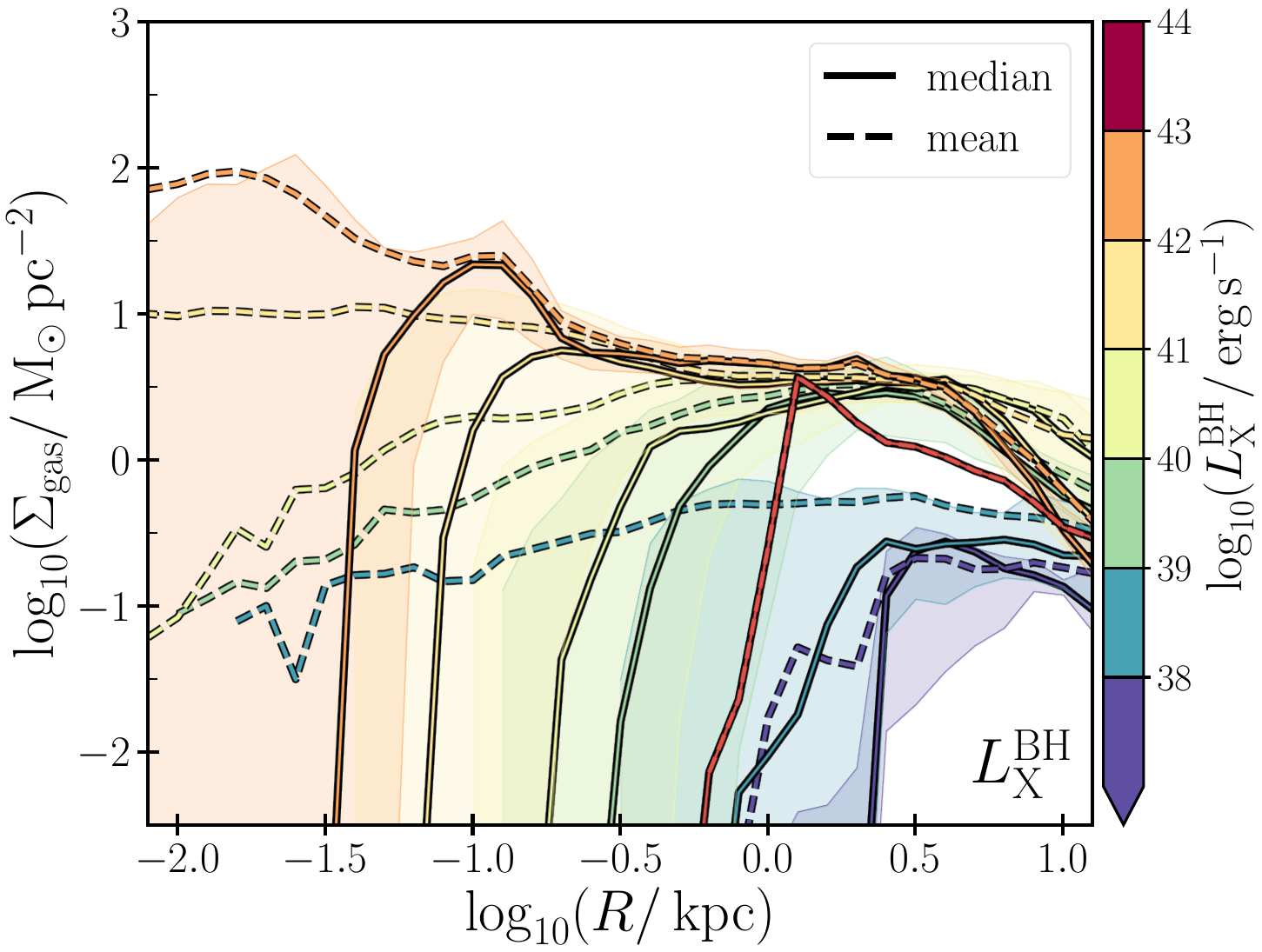}
\caption{Gas mass surface density ($\Sigma_{\rm gas}$) as a function of distance to the BH for different bins in the inflow-based luminosity proxy $L_{\rm X}^{\rm AD}$ (left) or the BH accretion-based X-ray luminosity $L_{\rm X}^{\rm BH}$ (right). 
Colours indicate the range of $L_{\rm X}$ in each bin, increasing from purple to red. Solid lines represent median radial profiles, with shaded regions enclosing the 25$^{\rm th}$-75$^{\rm th}$ percentiles, and dashed lines show the average radial profile in each bin. 
Galaxies with higher $L_{\rm X}^{\rm AD}$ (left panel) have consistently steeper radial density profiles, with increasing $\Sigma_{\rm gas}$ closer to the BH corresponding to stronger gas inflows down to the accretion disc, as expected for a predominant AGN fuelling phase.
Galaxies with higher instantaneous BH accretion rate (higher $L_{\rm X}^{\rm BH}$ in the right panel) often reflect a predominant AGN fuelling phase as well, but simulations also capture feedback-dominated phases with luminous AGN ($L_{\rm X}^{\rm BH}=10^{43-44}\,{\rm erg}\,{\rm s}^{-1}$; red) sustained by the accretion disc reservoir while AGN winds evacuate gas from the nuclear region.}
\label{fig:LXbins} 
\end{figure*}

Figure~\ref{fig:m12f_MandM} further explores BH self-regulation cycles by presenting the $\dot{M}_{\rm AD}$--$\dot{M}_{\rm BH}$ diagram for simulation \simf, covering the same time window as Figure~\ref{fig:all_timeplots}. We show \simf\ because it provides the clearest visualization of the recurrent loop-like behaviour over the full 1\,Gyr interval. The other galaxies exhibit analogous cycles, but over this long time window their trajectories overlap more strongly in the $\dot{M}_{\rm AD}$--$\dot{M}_{\rm BH}$ plane, so the individual loops are less easily separated. Both accretion rate histories are smoothed using a 1\,Myr running average (solid line) while points correspond to individual time steps, and the colour scale represents time evolution (from purple to yellow) over the 1\,Gyr period.
Grey dashed lines show different values of $\xi \equiv \dot{M}_{\rm BH}/ \dot{M}_{\rm AD}$, which measures the instantaneous response of the BH accretion rate to the rate at which gas is supplied to the unresolved accretion disc.\footnote{This quantity is distinct from the Eddington ratio, $\lambda_{\rm Edd}\equiv \dot{M}_{\rm BH}/\dot{M}_{\rm Edd}$, which compares the BH accretion rate to the radiation-pressure-limited growth rate of the BH. Instead, $\xi$ compares the BH accretion rate to the contemporaneous supply rate onto the subgrid disc. Observationally, $\dot{M}_{\rm BH}$ can be inferred indirectly from AGN luminosity, while $\dot{M}_{\rm AD}$ would require model-dependent estimates of nuclear inflow from resolved gas kinematics and gas densities. Therefore, $\xi$ itself is not directly measurable, although its time-averaged behaviour may be constrained indirectly through comparisons between AGN luminosity and nuclear gas supply.}
Specifically, we plot $\xi = 10,\, 1,\, 0.1,\, 0.01$, corresponding to BH accretion rates ranging from 10 times larger than the disc feeding rate down to only 1\% of $\dot{M}_{\rm AD}$.

The region of the $\dot{M}_{\rm AD}$--$\dot{M}_{\rm BH}$ diagram with $\xi > 1$ (upper left) corresponds to phases where the BH is accreting from a pre-existing gas disc reservoir that is not being efficiently replenished from larger scales, while the $\xi \ll 1$ region (lower right) corresponds to periods when substantial inflow rates increase the mass of the accretion disc before the subsequent increase in BH accretion rate establishes self-regulation. 
The overall trajectory of the BH in the $\dot{M}_{\rm AD}$--$\dot{M}_{\rm BH}$ diagram shows recurrent loop-like excursions that provide a phase-space view of the self-regulated AGN fuelling and feedback cycles. Similar behaviour is present in the other galaxies, but there the superposition of multiple cycles over the full 1\,Gyr interval makes the loop structure less visually distinct.

During each cycle, increasing disc inflow drives an increase in BH accretion, triggering feedback that disrupts the inflowing gas, reducing both $\dot{M}_{\rm AD}$ and subsequently $\dot{M}_{\rm BH}$. As cold gas replenishes the disc from larger scales, accretion resumes, completing the loop. For the representative loop-like cycles shown here, the durations span from a few $\times$10\,Myr to $\sim$150\,Myr, with typical values $\sim$100\,Myr.
This recurrent loop-like trajectory is shaped by the subgrid $\alpha$-disc model, which smooths accretion rate variations over the characteristic timescale $t_{\rm acc}$, which depends on the total masses of the BH and its accretion disc rather than directly on either $\dot{M}_{\rm AD}$ or $\dot{M}_{\rm BH}$ (Equation~\ref{eq:tacc}) and moderates the short-term response of the BH. Thus, the loops reflect both the intrinsic variability of inflow and feedback and the finite response time of the unresolved accretion-disc reservoir. In this sense, departures of $\xi$ from unity trace the lag between gas supply to the subgrid disc and accretion onto the BH.

Overall, Figures~\ref{fig:all_timeplots} and~\ref{fig:m12f_MandM} show that AGN activity is intrinsically episodic, with alternating periods of feeding and feedback modulated by the response of the accretion disc to the inflow rate from larger scales and the coupling efficiency of AGN winds with the surrounding ISM on $\sim$10--100\,pc scales.

\section{Gas concentration -- $L_{\rm X}$ relation} \label{sec:correlation}
Motivated by observations of local Seyfert galaxies \citep{Garcia-Burillo2021,Garcia-Burillo2024}, Figure~\ref{fig:m12i_multipanel} explores the connection between gas concentration and the morphology of dense gas in the nuclear region of a representative simulation (\simi). 
The panels on the left illustrate the time evolution of gas concentration (top; $\Sigma_{\rm inner}/\Sigma_{\rm outer}$) and BH kernel radius (bottom) for different definitions. Following \citet{Garcia-Burillo2024}, we compute gas concentration as the ratio of gas surface density within the central 50\,pc to that within 200\,pc (blue x markers), where we consider all non-wind gas particles rather than an explicitly selected molecular or CO-traced component.
For comparison, we also estimate gas concentration based on variable BH kernels (plotted in the bottom panel) with radii defined to enclose the nearest 32 gas elements ($R_{32}$) and the nearest 256 gas elements ($R_{256}$). This provides an alternative definition designed to characterise variations in the nuclear gas concentration while allowing the radial apertures to adjust dynamically to the local gas distribution (red triangle markers).\footnote{$R_{256}$ corresponds to the BH kernel radius, while $R_{32}$ defines an inner aperture using the same adaptive neighbour construction. We retain this adaptive concentration diagnostic for several reasons. First, there are many snapshots in which gas is present within 200\,pc but no gas resolution element lies within the central 50\,pc, such that the fixed aperture concentration can only be represented by an upper limit. The adaptive apertures continue to characterise the distribution of the resolved nuclear gas during these states of strong depletion. Second, $R_{256}$ is the same BH kernel scale used to evaluate $\Sigma_{256}$ in Figure~\ref{fig:all_timeplots} and the gas properties entering the BH accretion model, providing an internally consistent measure of the gas distribution around the BH. 
Since mechanical feedback is implemented through newly spawned wind particles rather than deposition into pre-existing gas within the BH kernel (\S\ref{subsec:bhphysics}), 
$R_{256}$ does not define the gas elements receiving mechanical feedback and therefore it does not represent a prescribed feedback cavity radius.
Finally, although the fixed 50\,pc and 200\,pc apertures are essential for a direct comparison with the observational concentration index, they are observationally motivated scales rather than unique fundamental scales for BH fuelling or feedback. We therefore use $\Sigma_{R_{32}}/\Sigma_{R_{256}}$ as a complementary, resolution-dependent  diagnostic of the simulated gas distribution. Because these radii depend on the resolved gas sampling, we do not use the adaptive concentration for direct quantitative comparison with the observed concentration indices.} 
Open markers for the $\Sigma_{\rm 50\,pc}/\Sigma_{\rm 200\,pc}$ gas concentration definition indicate instances where gas is present within 200\,pc but absent within 50\,pc due to resolution limitations, which motivate the dynamical definition based on $\Sigma_{R_{32}}/\Sigma_{R_{256}}$. In such cases, we assume an upper limit of one gas element (with mass $m_{\rm b} = 7000\,\Msun$) at $R<50$\,pc to estimate $\Sigma_{\rm 50\,pc}$.
The large fluctuations in gas concentration and enclosed radii $R_{32}$ and $R_{256}$ reflect dynamic processes of gas inflow, outflow, and redistribution within the nuclear region.

The right panels of Figure~\ref{fig:m12i_multipanel} show a series of face-on projected gas mass surface density maps for the central 500\,pc region, corresponding to representative times highlighted as markers with black outline in the left panels. The inner and outer gray dashed circles indicate the central 50\,pc and 200\,pc regions, respectively, which are used to calculate the central gas concentration $\Sigma_{\rm 50\,pc}/\Sigma_{\rm 200\,pc}$. Brighter regions represent areas with higher gas density, whereas darker regions indicate gas-poor areas. Visually, we observe a variety of morphological features quickly developing over time (timescales as short as $\sim$30\,Myr), with frequent instances of nuclear spirals, accumulation of central gas reservoirs, and cavity formation. 
Selected snapshots include some of the highest and lowest values of $\Sigma_{\rm 50\,pc}/\Sigma_{\rm 200\,pc}$ identified in the simulation, which we analyse below in the context of BH accretion and feedback.

The changes in gas concentration and morphological evolution of the nuclear gas density distribution reflect the AGN fuelling and feedback cycles analysed in \S\ref{sec:gas_properties}. Visual evidence of increasing gas concentration followed by the formation of gas-poor cavities directly correspond to the strong variations in $\log_{10}(\Sigma_{\rm 50\,pc}/\Sigma_{\rm 200\,pc})$ shown in the top-left panel of Figure~\ref{fig:m12i_multipanel}. 
For example, at $t = 13.051$\,Gyr we see a central cavity with size $r \sim50$\,pc surrounded by a nuclear gas disc with prominent spiral arms on $\sim$100\,pc scales, corresponding to very low concentration index ${\rm CCI}\equiv{\rm log}_{10}(\Sigma_{\rm 50\,pc}/\,\Sigma_{\rm 200\,pc}) < -1.8$. In the following snapshot ($\sim$30\,Myr later), the nuclear gas disc has replenished the central 50\,pc, reaching high gas concentration CCI\,$\approx 0.75$, with the subsequent opening of a central cavity with very asymmetric distribution at $t = 13.118$\,Gyr and the consequent drop in concentration index. Similar gas concentration cycles occur over time, such as at $t =13.282$\,Gyr where another strong peak in concentration is followed by a prominent gas cavity.
Gas concentration measurements based on $\Sigma_{R_{32}}/\Sigma_{R_{256}}$ generally trace the changes measured on 50--200\,pc scales, although with quantitative differences as expected.  
Morphological changes in the gas distribution are thus well aligned with the quantitative fluctuations in the gas concentration index, with repeated cycles of gas accumulation and feedback-driven removal representing a key signature of self-regulated BH growth.

In order to compare simulations to the observed CCI--$L_{\rm X}$ relation, we estimate the hard X-ray luminosity in the 2--10\,${\rm keV}$ band using the bolometric correction prescription from \citet{Shen2020}. This method is based on the double power-law model introduced by \citet{Hopkins2007}, which fits the dependence of bolometric corrections on AGN bolometric luminosity:
\begin{equation}\label{eq:Lx_equation}
    \frac{L_{\mathrm{bol}}}{L_{\mathrm{band}}} = c_1 \left( \frac{L_{\mathrm{bol}}}{10^{10}\,{\rm L}_\odot} \right)^{k_1} + c_2 \left( \frac{L_{\mathrm{bol}}}{10^{10}\,{\rm L}_\odot} \right)^{k_2}.
\end{equation}

We note that this formulation does not account for extinction, which may lead to lower observed X-ray luminosities in heavily obscured systems.\footnote{The $L_{\rm X}$ values inferred in this way should be interpreted as intrinsic, bolometric-correction-based 2--10\,keV luminosities rather than forward-modeled observed X-ray fluxes. This is broadly consistent with comparing to absorption-corrected hard X-ray luminosities in the observational samples, but we do not model line-of-sight obscuration explicitly. Residual uncertainties in the absorption correction, especially for heavily obscured or Compton-thick systems, could shift some observed systems to lower intrinsic or apparent $L_{\rm X}$. This effect mainly affects the horizontal placement of points in the CCI--$L_{\rm X}$ plane and does not directly change the gas concentration itself.}
The best-fitting parameters for various bands are given in Table 1 of \citet{Shen2020}. For this paper, we consider the hard X-ray band corresponding to the integrated luminosity in 2--10\,keV, with bolometric correction parameters $[c_1, k_1, c_2, k_2]=[4.073,-0.026,12.60,0.278]$. We estimate bolometric luminosities based on both the inflow rate down to the accretion disc ($\Mdot_{\rm AD}$) and based on the BH accretion rate ($\MdotBH$) using $L_{\rm bol}=\epsilon_{\rm rad} \Mdot c^{2}$, where we adopt $\epsilon_{\rm rad} = 0.1$ for consistency with the actual radiative efficiency adopted in the FIRE-3 simulations. 

The radiative luminosity of the AGN is associated with accretion through the innermost flow onto the BH, such that $L_{\rm X}^{\rm BH}$, derived from $\dot{M}_{\rm BH}$, represents the physically motivated X-ray luminosity prediction of the two-stage accretion model used here. We nevertheless also define $L_{\rm X}^{\rm AD}$ by applying the same conversion to $\dot{M}_{\rm AD}$, which should be interpreted as an inflow-based luminosity proxy rather than as radiation physically emitted at the resolved accretion disc feeding scale.
We implement this second estimate because the mapping between resolved nuclear inflow and BH accretion remains highly uncertain in galaxy formation simulations. In models without an explicit intermediate accretion disc reservoir, the subgrid inflow rate estimated from resolved scales has commonly been identified directly with the BH accretion rate. Comparing $L_{\rm X}^{\rm AD}$ and $L_{\rm X}^{\rm BH}$ therefore allows us to isolate the effect of introducing the finite subgrid accretion disc reservoir and its associated response time on the predicted concentration--luminosity relation.

Figure~\ref{fig:comparison_to_obs} shows the gas concentration as a function of X-ray luminosity for the last Gyr of our four simulated galaxies, compared to observations (indicated by different markers, as labeled).
The top row uses the definition of gas concentration based on the 50\,pc and 200\,pc regions. Red arrows indicate upper limits, representing instances where there is no gas in the central 50\,pc. The bottom row employs our alternative definition of gas concentration using the variable radial apertures $R_{32}$ and $R_{256}$.
The left column corresponds to the inflow-based luminosity proxy $L_{\rm X}^{\rm AD}$, while the right column shows the X-ray luminosity $L_{\rm X}^{\rm BH}$ inferred from the BH accretion rate.
The marginal histograms show the corresponding one-dimensional distributions of $L_{\rm X}$ (top) and gas concentration (right) for the simulated and observed points in each panel. These histograms are intended to visualize the projected distributions along each axis, rather than to compare occurrence rates directly, since the simulated tracks and observational samples have different sampling and selection functions.
The corresponding time-domain evolution of these luminosity and concentration diagnostics is shown in Figure~\ref{fig:all_CCI_timeplots}, which illustrates how the fixed-aperture and dynamically defined concentration measures can differ across the four simulations.
White circles represent AGN targets from the GATOS-NUGA survey \citep[with the corresponding black arrows denoting upper limits;][]{Garcia-Burillo2024} and white triangles show galaxies from the millimetre-Wave Interferometric Survey of Dark Object Masses (WISDOM) survey \citep{Elford2024}, shown in the upper panels. These observational points should be interpreted with their sample definitions in mind. 
The GATOS-NUGA comparison is based on nearby late-type Seyfert galaxies with high-resolution CO observations of their circumnuclear molecular gas, whereas WISDOM targets nearby galaxy centres for dynamical and molecular gas studies and includes a broader host population, including many early-type galaxies and a wider range of nuclear activity levels.
Therefore, the comparison samples are not designed to be volume-complete at fixed $L_{\rm X}$ or complete in CCI. In particular, inclusion in this diagram requires resolved CO measurements on the $\sim$50--200\,pc scales used to define the concentration index.
The vertical black dashed line at ${\rm log}_{10}(L_{\rm X}/\,{\rm erg}\,{\rm s}^{-1})=41.5$ separates the build-up phase (left region) from the feedback-dominated phase (right region) as defined in \citet{Garcia-Burillo2024}, with their two-branch linear fit shown as blue and red lines for the build-up and feedback phases, respectively. 

The spatial scales probed by the observations and simulations are comparable but not identical. The observed CO maps resolve the central regions on scales of order $\sim$10\,pc, sufficient to measure average surface densities within 50\,pc and 200\,pc apertures. Since this spatial resolution is substantially smaller than the apertures used to define the gas concentration, we do not expect beam smearing to significantly bias the observed $\Sigma_{\rm 50\,pc}/\Sigma_{\rm 200\,pc}$ measurements or to account for the differences with simulations discussed below.
In the simulations, the minimum adaptive gravitational softening for gas is $\sim$0.14\,pc, well below the fixed apertures used here, and the baryonic mass resolution is $m_{\rm b}=7000\,\Msun$. However, our CCI is measured directly from the simulated gas distribution, whereas the observed CCI is inferred from
CO emission tracing the molecular gas. Differences in the gas phase being traced, molecular-line sensitivity, and the conversion between CO emission and molecular gas surface density therefore represent more relevant limitations of the present comparison than the observational beam size.

Starting from the top left panel, the inflow-based luminosity proxy $L_{\rm X}^{\rm AD}$ spans ${\rm log}_{10}(L_{\rm X}^{\rm AD}/\,{\rm erg}\,{\rm s}^{-1}) = 38$--43.5, broadly overlapping the numerical range of X-ray luminosities in the observational samples. We emphasize that $L_{\rm X}^{\rm AD}$ is not interpreted as radiation physically produced at the accretion disc feeding scale. Instead, this panel represents the luminosity that would be inferred if the resolved inflow rate were identified directly with the BH accretion rate, allowing us to assess the effect of the intermediate subgrid accretion disc reservoir. We also find that simulations exhibit a broader distribution in gas concentration compared to observations, as also seen in the right-hand marginal histogram for the fixed-aperture concentration. Observed galaxies rarely reach nuclear gas concentrations below ${\rm log}_{10}(\Sigma_{\rm 50\,pc}/\Sigma_{\rm 200\,pc})\lesssim -0.5$, with such low values corresponding to the most extreme nuclear gas deficits. By contrast, our simulated sample extends well below this lower observed range, reaching ${\rm log}_{10}(\Sigma_{\rm 50\,pc}/\Sigma_{\rm 200\,pc})\approx -2$, while spanning an overall range from -2 to 1. We do not interpret the absence of observed counterparts with CCI $<-0.5$ as a direct completeness statement, since the observed samples require resolved CO detections and are subject to molecular-line sensitivity and tracer-selection effects. The simulations produce very low-concentration nuclear states that are thus either absent, rare, short-lived, or difficult to recover in the current observational comparison samples.

Interestingly, we observe a sharp transition near the threshold separating the AGN build-up and feedback phases ($L_{\rm X}^{\rm AD} > 10^{41.5}\,{\rm erg}\,{\rm s}^{-1}$), where simulated galaxies span a wide range of gas concentrations while the central BH displays a narrower range of X-ray luminosities, indicating a plausible threshold $L_{\rm X}$ for efficient AGN feedback as suggested by observations. The top marginal histograms further emphasize that the simulated and observed samples overlap broadly in $L_{\rm X}$, while the main panels and right-hand histograms show that the simulations extend to lower gas concentrations. In particular, individual simulated galaxies need not evolve smoothly along the observed build-up and feedback branches. Instead, the simulated points suggest that changes in gas concentration can be larger than changes in $L_{\rm X}$ over short timescales. On the other hand, simulations populate the upper-left quadrant of the CCI--$L_{\rm X}$ plane less frequently than observations, indicating that gas-rich, low-$L_{\rm X}$ states are less common in the simulated tracks shown here. Because the observational samples are not complete at fixed $L_{\rm X}$ or fixed CCI, this difference should not be interpreted as a direct difference in occurrence rates. If the observed upper-left population is representative, then better agreement would require simulated systems to spend more time in phases where nuclear gas has built up before the X-ray luminosity rises, or would require additional variability/obscuration effects that move gas-rich systems toward lower observed $L_{\rm X}$.

\begin{figure*}
\includegraphics[width = \columnwidth]{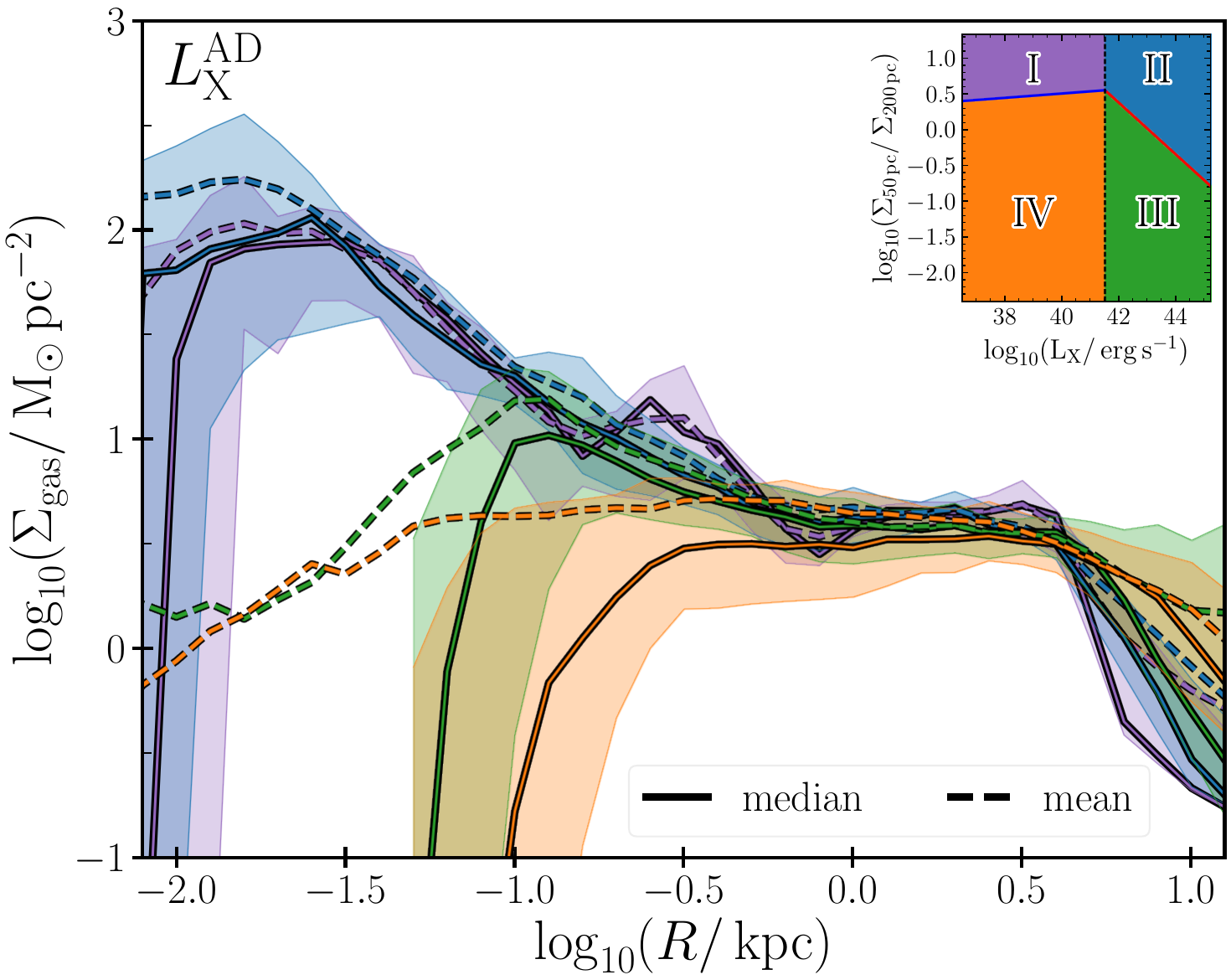}
\hspace{2.5mm}
\includegraphics[width = \columnwidth]{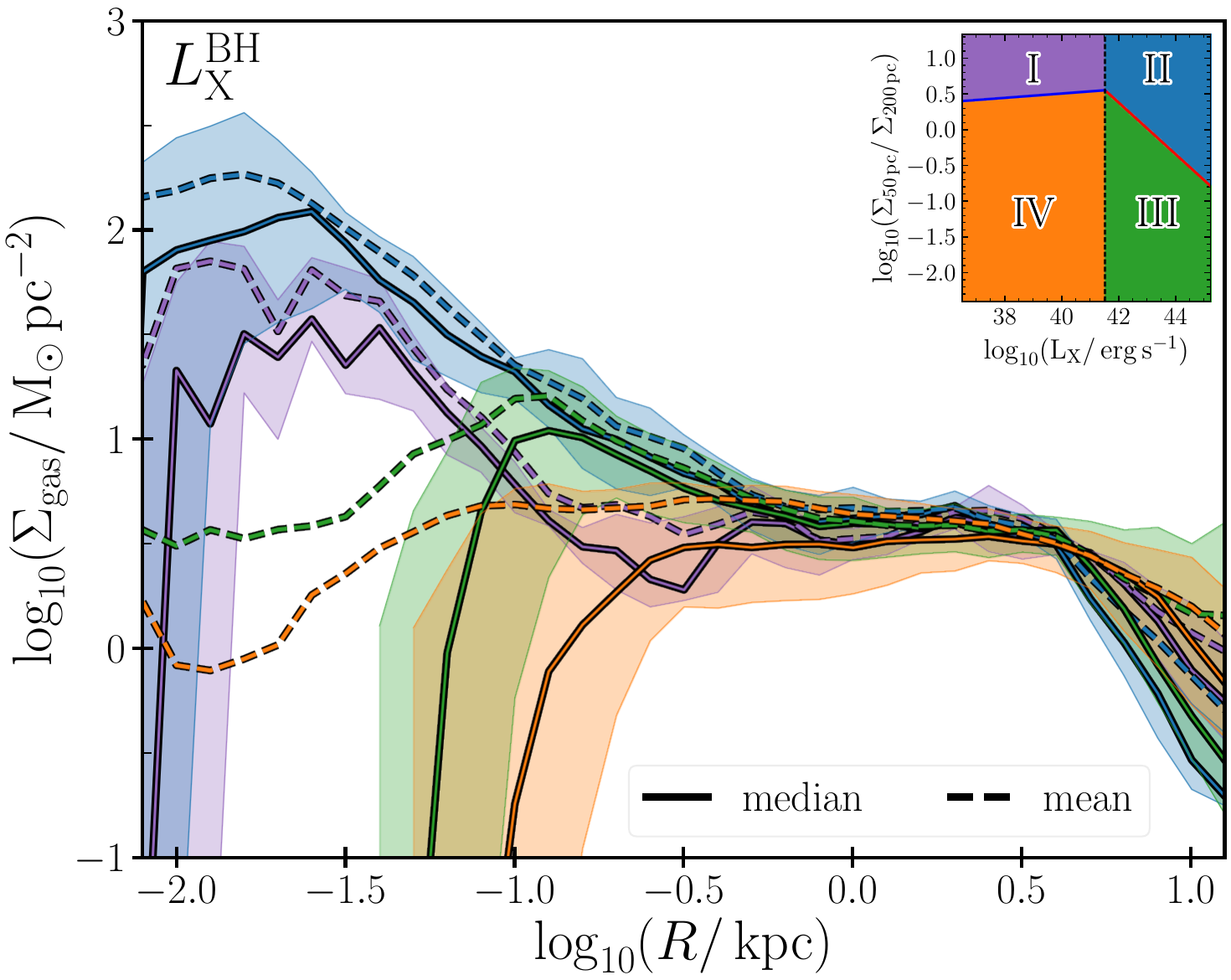}
\caption{Similar to Figure~\ref{fig:LXbins} but binning gas density profiles according to four different quadrants in the gas concentration--$L_{\rm X}$ diagram motivated by observations of Seyfert galaxies in the GATOS-NUGA surveys \citep{Garcia-Burillo2024}, as shown in Figure~\ref{fig:comparison_to_obs} and indicated by the inset panels here. 
$L_{\rm X}$ is estimated based on the inflow rate down to the accretion disc ($L_{\rm X}^{\rm AD}$; left) or based on the accretion rate onto the BH ($L_{\rm X}^{\rm BH}$; right).
Solid lines represent median radial profiles, with shaded regions enclosing the 25$^{\rm th}$-75$^{\rm th}$ percentiles, and dashed lines show the average radial profile in each bin.
Higher luminosity AGN in simulated galaxies are found in both fuelling-dominated (Quadrant II; blue) and feedback-dominated (Quadrant III; green) phases, corresponding to centrally-peaked $\Sigma_{\rm gas}$ profiles and evacuated central cavities, respectively. Lower luminosity AGN often show nuclear gas deficits driven by previous feedback events (Quadrant IV; orange) but can also show centrally rising $\Sigma_{\rm gas}$ profiles indicating evolution toward the next re-fuelling event (Quadrant I; purple).
}
\label{fig:quadrant_bins} 
\end{figure*}

The top right panel of Figure~\ref{fig:comparison_to_obs} shows the corresponding result after the intermediate accretion disc reservoir is included, using $L_{\rm X}^{\rm BH}$ derived from the actual BH accretion rate, which exhibits less variation and is confined to the narrower range of ${\rm log}_{10}(L_{\rm X}^{\rm BH}/\,{\rm erg}\,{\rm s}^{-1}) \approx 40.5$--43. This reduced dynamic range in X-ray luminosity reflects the structure of the BH accretion model used in the simulations, demonstrating that introducing the finite subgrid accretion disc reservoir substantially modifies the predicted AGN luminosity distribution. In our two-stage model, gas has to first feed the accretion disc from the nuclear gas reservoir and then accrete from the disc down to the BH on a characteristic accretion timescale (Equation~\ref{eq:tacc}). Because $\dot{M}_{\rm BH}$ is set by the mass stored in the subgrid accretion disc, rather than by the instantaneous inflow rate onto the disc, the accretion disc acts as a reservoir that smooths short-timescale fluctuations in $\dot{M}_{\rm AD}$. As a result, $L_{\rm X}^{\rm BH}$ varies less strongly than $L_{\rm X}^{\rm AD}$, and individual simulated galaxies tend to shift more strongly in gas concentration than in $L_{\rm X}^{\rm BH}$. Their motion in this plane is therefore often closer to vertical than to a smooth progression along the observed branches. Feedback from the BH can then persist even as the central gas reservoir begins to deplete, producing a more stable and prolonged impact on the surrounding gas distribution. This may explain the sharp decrease in nuclear gas concentration seen in simulations at roughly constant $L_{\rm X}^{\rm BH}$.
 
Finally, the bottom panels illustrate the dependence of the concentration--$L_{\rm X}$ relation on the definition of gas concentration, which in this case is based on the ratio of gas surface densities within radii defined to enclose a fixed number of gas resolution elements (32 and 256 for $R_{32}$ and $R_{256}$, respectively). Since these radii dynamically change depending on how gas is distributed, we can always compute the ratio $\Sigma_{R_{32}}/\Sigma_{R_{256}}$ even in the presence of large cavities.
As a result, the number of simulated data points in the concentration--$L_{\rm X}$ diagram increases without relying on upper limits. Because this concentration measure is tied to the resolved gas sampling rather than fixed physical scales, we do not compare its absolute values directly with the observed concentration indices. Instead, the bottom panels provide a simulation to simulation comparison of how the inferred concentration-luminosity behaviour changes when the nuclear gas distribution is characterised using adapative rather than fixed apertures. The simulations show a narrower range of concentration with the adaptive defintion, ${\rm log}_{10}(\Sigma_{R_{32}}/\Sigma_{R_{256}}) \approx -1.5$--0.5, while retaining the same broad episodes of nuclear gas accumulation and depletion identified using the fixed apertures.

Overall, simulations cover a broadly similar range of parameter space as observations in the concentration--$L_{\rm X}$ diagram but do not show a clear correlation between gas concentration and X-ray luminosity.
This comparison should be interpreted as a qualitative test of whether the simulations occupy the observed CCI--$L_{\rm X}$ plane, rather than as a direct one-to-one prediction of observed number densities. The extended population of very low-CCI simulated states represents a clear discrepancy with observations. 
If these states remain after constructing observationally motivated synthetic CO measurements with appropriate sensitivity limits and applying a consistent treatment of X-ray obscuration, this would suggest that the simulations produce central gas depletion that is too deep, too frequent, or too long-lived. Conversely, if mock observations move these points toward the observed lower envelope, then part of the difference would arise from how intrinsic gas distributions are translated into observed CO-based concentration indices.

\section{Nuclear gas density profiles} \label{sec:density_profiles}

Figure~\ref{fig:LXbins} presents radial profiles of gas mass surface density for our simulated galaxies in different bins of AGN X-ray luminosity, where each bin includes all galaxies at all times falling within the respective luminosity range. The left panel bins radial profiles based on $L_{\rm X}^{\rm AD}$ while the right panel corresponds to bins in $L_{\rm X}^{\rm BH}$. In both panels, lower luminosity bins are represented in blue hues, transitioning to red hues for higher luminosities. We use logarithmically-spaced intervals of 1 dex up to ${\rm log}_{10}(L_{\rm X}/\,{\rm erg}\,{\rm s}^{-1})=44$, and galaxies with ${\rm log}_{10}(L_{\rm X}/\,{\rm erg}\,{\rm s}^{-1})\leq 38$ are included in the lowest luminosity bin. 
Solid lines depict median radial profiles, dashed lines represent geometric average profiles, and  shaded regions encompass the 25$^{\rm th}$ to 75$^{\rm th}$ percentiles of the distribution to characterize the typical nuclear gas surface densities associated with each BH accretion state.

The left panel shows a clear correlation between the inflow rate down to the accretion disc (parameterized by $L_{\rm X}^{\rm AD}$) and the gas density distribution, with higher X-ray luminosities corresponding to more centrally peaked gas radial profiles (reaching $\Sigma_{\rm gas} \sim 300\,\Msun\,{\rm pc}^{-2}$ at $R\sim10$\,pc for $L_{\rm X}^{\rm AD} > 10^{43}\,{\rm erg}\,{\rm s}^{-1}$) and lower X-ray luminosities corresponding to progressively larger central cavities (reaching $R_{\rm cavity} \gtrsim 2$\,kpc at $L_{\rm X}^{\rm AD} \leq 10^{38}\,{\rm erg}\,{\rm s}^{-1}$). This trend suggests that higher central gas concentrations enable more efficient fuelling of the accretion disc, resulting in increased X-ray luminosities in a fuelling-dominated regime. 
The right panel shows a qualitatively similar trend for the BH accretion rate (parameterized by $L_{\rm X}^{\rm BH}$), with more centrally peaked $\Sigma_{\rm gas}$ profiles at higher X-ray luminosity. However, in this case, the highest luminosity bin appears to be reverting the trend, with a prominent $\sim$kpc-scale cavity forming at $L_{\rm X}^{\rm BH} > 10^{43}\,{\rm erg}\,{\rm s}^{-1}$. 
While infrequent (there is only one snapshot identified at  $L_{\rm X}^{\rm BH} > 10^{43}\,{\rm erg}\,{\rm s}^{-1}$), this indicates that our simulations are capable of reproducing feedback-dominated phases where a luminous AGN is powered by a pre-existing accretion disc reservoir while AGN winds evacuate gas from the nuclear region.

In Figure~\ref{fig:quadrant_bins}, we further explore how nuclear gas density profiles vary with AGN luminosity by organizing the simulated data into four representative quadrants defined in the concentration--$L_{\rm X}$ diagram, as shown in the inset panels (see also Figure~\ref{fig:comparison_to_obs} and~\ref{fig:comparison_to_obs_alltimes}). As in Figure~\ref{fig:LXbins}, this binning is based on the full ensemble of snapshots from all four galaxies over all available times. These quadrants divide the parameter space of gas concentration--$L_{\rm X}$ according to the threshold X-ray luminosity $L_{\rm X} = 10^{41.5}\,{\rm erg}\,{\rm s}^{-1}$ identified in \citet{Garcia-Burillo2024} and their two-branch fit for observed concentration index in Seyfert galaxies as a function of X-ray luminosity.
Here we compute gas concentration as the ratio ${\rm log}_{10}(\Sigma_{\rm 50\,pc}/\Sigma_{\rm 200\,pc})$, rather than using our alternative definition that uses $R_{32}$ and $R_{256}$, so that the binning more directly matches the physical scales probed by the observations.
We estimate X-ray luminosity based on either the inflow rate down to the accretion disc ($L_{\rm X}^{\rm AD}$; left panel) or the BH accretion rate ($L_{\rm X}^{\rm BH}$; right panel).
For each quadrant, we compute median (solid lines), mean (dashed lines), and 25$^{\rm th}$–75$^{\rm th}$ percentiles (shaded regions) over all the contributing gas mass surface density radial profiles from snapshots assigned to the corresponding range of gas concentration and X-ray luminosity. Snapshots with no gas within 50\,pc are assigned to quadrants using the corresponding concentration upper limit, but their radial profiles are otherwise left unchanged.
Colour coding corresponds to the four quadrants: low-luminosity/high-concentration (Quadrant I; purple), high-luminosity/high-concentration (Quadrant II; blue), high-luminosity/low-concentration (Quadrant III; green), and low-luminosity/low-concentration (Quadrant IV; orange). 

These quadrant-based categories can be interpreted as potential sequential stages in a regulated cycle of AGN fuelling and feedback, broadly consistent with the qualitative evolutionary scenario proposed by \citet[][\S8]{Garcia-Burillo2024}. At the ensemble level, individual galaxies need not traverse all four quadrants in sequence, but the quadrant populations are consistent with such a cycle. A recent nuclear fuelling event can increase the gas concentration on $\sim$50--200\,pc scales while the AGN is still in a previous low-luminosity state, corresponding, when populated, to Quadrant I. Galaxies can then transition into Quadrant II after a dynamical delay \citep{Hopkins2012_DynamicalDelay,Angles-Alcazar2021} allowing nuclear gas to lose enough angular momentum to trigger vigorous BH accretion, simultaneously displaying high nuclear gas concentration and elevated $L_{\rm X}$.
The higher BH accretion rate increases the impact of AGN feedback, which can subsequently evacuate gas from the central region and decrease the gas concentration while the accretion disc reservoir continues to power high $L_{\rm X}$, transitioning galaxies into Quadrant III. As nuclear fuelling diminishes and the pre-existing accretion disc is consumed, the AGN luminosity drops and the galaxy moves into Quadrant IV. With time, renewed gas inflow from larger scales can rebuild the nuclear reservoir and increase gas concentration, which is consistent with re-populating Quadrant I in the ensemble view.

This quadrant-based analysis offers physical insight into the coupling between gas inflow, BH accretion, and feedback-driven depletion. In both $L_{\rm X}^{\rm AD}$ and $L_{\rm X}^{\rm BH}$ panels, we find that the more centrally concentrated quadrants (I and II) exhibit steep central gas gradients, while Quadrants III and IV are characterised by flatter profiles and central cavities, confirming the concentration index as an effective parameterization of the diversity of gas density profiles \citep{Garcia-Burillo2024}.
While higher central gas concentration tends to correlate with higher $L_{\rm X}$ in simulated galaxies (Figure~\ref{fig:LXbins}), we find that luminous AGN can populate both Quadrant II, reflecting a fuelling-dominated phase with high gas concentration (blue), and Quadrant III, reflecting a feedback-dominated phase with central cavities (green).
Similarly, lower-luminosity AGN are found primarily in Quadrant IV, corresponding to nuclear gas deficits after previous feedback episodes (orange), while the more sparsely populated Quadrant I is consistent with lower-luminosity states that retain relatively high central gas concentration (purple).
Interestingly, while these trends are qualitatively similar for both accretion proxies, the difference in central gas densities between high- and low-luminosity states is more pronounced when using $L_{\rm X}^{\rm BH}$ compared to $L_{\rm X}^{\rm AD}$, suggesting that the BH accretion model smooths variability through the imposed accretion timescale, thereby sustaining feedback over longer periods even as gas is depleted.

\section{Discussion} \label{sec:discussion}

Modern cosmological hydrodynamic simulations show that SMBH growth and feedback can greatly influence the evolution of galaxies and large-scale structure \citep{Somerville&Dave2015,Crain_vandeVoort_review2023}, playing a key role suppressing the growth of massive galaxies \citep{Choi2018,Grayson2026}, impacting the properties of gas across scales \citep{Zinger2020,Tillman2023}, and altering the overall distribution of matter in the universe \citep{vanDaalen2020,Delgado2023,Gebhardt2024,Gebhardt2026,Bigwood2025}. Given the limited resolution, simulations must treat BH physics at a subgrid level, often relying on phenomenological models and extensive tuning of free parameters. In most models, key parameters such as BH accretion and feedback efficiencies are tuned to match population-level statistics such as galaxy stellar mass functions \citep{Weinberger2017,Dave2019} and BH--galaxy scaling relations \citep{Thomas2019,Habouzit2021,Shankar2025}. While simulations can then be tested against independent observables, such as CGM properties \citep{Oppenheimer2020,Lau2025} and AGN luminosity functions \citep{Habouzit2022}, global observables are often not sufficient to discriminate between models. The development of higher-resolution cosmological zoom-in simulations with more detailed ISM physics is now enabling the possibility of implementing and testing BH accretion and feedback models on $\sim$10--100\,pc scales \citep{Hopkins2023_fire3,Goddard2025}, leveraging the wealth of spatially-resolved, multi-wavelength data available in the nearby universe \citep{Husemann2022,Harrison2024}.

In this work we have focused on high-resolution observations of nearby Seyfert galaxies to constrain BH physics in the FIRE-3 simulations. In particular, the GATOS survey \citep{Garcia-Burillo2021,Garcia-Burillo2024} finds that higher-luminosity Seyferts show increasing central molecular gas deficits within $\sim$50\,pc, suggesting evolution along an ``AGN feedback branch,'' while lower-luminosity systems tend to retain centrally peaked gas distributions, suggesting nuclear refuelling along an ``AGN build-up branch.''
The WISDOM project \citep{Elford2024} finds no simple correlation between circumnuclear gas mass and AGN luminosity, possibly reflecting both sample diversity and the stochastic nature of nuclear activity. In particular, WISDOM includes many early-type galaxies (ellipticals and lenticulars), whereas the sample in GATOS is restricted to disc galaxies; this difference in host morphology may itself contribute to the different trends, since the secular processes that redistribute molecular gas in early-type and spiral galaxies are not expected to be the same. This is also relevant for our comparison, as the four simulated galaxies analysed here are disc galaxies by $z=0$ and GATOS is therefore a cleaner comparison sample. 

Our FIRE-3 cosmological zoom-in simulations of Milky Way-mass galaxies resolve a multiphase ISM and the coupling of AGN winds on $\lesssim$10\,pc scales \citep{Hopkins2023_fire3, Byrne2024}. These winds can excavate nuclear cavities on $\sim$10--500\,pc scales while preferentially propagating away from the dense disc along the polar direction, producing asymmetric, biconical structures similar to those found in idealized high-resolution models of AGN feedback \citep[e.g.][]{Hopkins2016,Torrey2020,Costa2020,Ward2024,Sivasankaran2025,Sivasankaran2026} and some cosmological zoom-in simulations \citep{Goddard2025}. At the same time, gas reaching the BH kernel is first transported from resolved nuclear scales down to a subgrid accretion disc reservoir, introducing a finite delay between resolved nuclear inflow and AGN luminosity. These two ingredients, resolved AGN wind-ISM coupling and the delayed response of the subgrid accretion disc, are key aspects of the model directly constrained by the connection of nuclear gas concentration and AGN luminosity identified in the GATOS survey.

Within this framework, our FIRE-3 simulations produce nuclear gas morphologies, global galaxy properties, and AGN luminosities that are broadly consistent with those observed in local Seyfert galaxies.   
However, the degree of apparent agreement with observations depends sensitively on how nuclear gas concentration and X-ray luminosity are measured. Using fixed physical apertures of 50\,pc--200\,pc allows for a more direct comparison to observations (relative to the alternative $R_{32}$--$R_{256}$ dynamical apertures) but can result in upper limits to the gas concentration index and produce nearly vertical excursions in the CCI--$L_{\rm X}$ plane that are not seen in observations. 
In addition, predicted BH accretion rates from the subgrid $\alpha$-disc yield a narrower range of variation in $L_{\rm X}$ compared to the inflow rate down to the disc as well as observations. These differences show that quantitative comparisons between simulations and data are sensitive to aperture choice and to the adopted proxy for AGN luminosity. 

\begin{figure}
\includegraphics[width = \columnwidth]{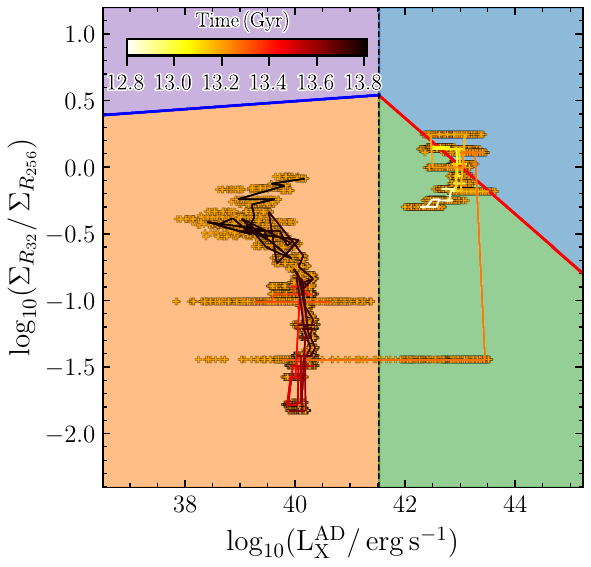}
\caption{
Gas concentration versus X-ray luminosity over the last 1\,Gyr of evolution for galaxy~\simb. 
BH-specific outputs provide high-cadence $L_{\rm X}^{\rm AD}$ measurements, which are assigned to the adaptive gas concentration $\Sigma_{R_{32}}/\Sigma_{R_{256}}$ computed for the closest data snapshot available. The line connecting data points indicates the trajectory of the galaxy, smoothed using cubic spline interpolation and colour-coded by simulation time. 
The galaxy begins near the high $L_{\rm X}$ branch of the broken power-law fit to GATOS sources \citep[blue/red solid lines;][]{Garcia-Burillo2024} before undergoing a strong AGN outflow-driven depletion phase that reduces the gas supply onto the accretion disc and moves the system to the lower-left quadrant. In this case, the galaxy remains in a low-luminosity/low-concentration state for several hundred Myr before gas gradually accumulates and refills the cavity.   This strong feedback event has long-lasting effects compared to the shorter timescale of the more typical accretion-feedback self-regulation cycles shown in Figure~\ref{fig:all_timeplots}.
}
\label{fig:timetrack_CCI_vs_Lx} 
\end{figure}

The simulations exhibit recurrent fuelling and depletion cycles on characteristic timescales $\sim$10--100\,Myr, similar to the variability timescales inferred for the GATOS sample, but they do not recover a clear gas concentration--$L_{\rm X}$ correlation as reported by \citet{Garcia-Burillo2024}. The important discrepancy is therefore not whether the simulations can produce nuclear cavities, as they are able to, but when these depleted states occur relative to luminous AGN phases. High-luminosity simulated AGN more frequently coincide with centrally concentrated gas than inferred from the GATOS ``feedback'' branch, suggesting that the timing between nuclear gas inflow, accretion disc response, and subsequent gas clearing is a key model uncertainty. Tracing individual self-regulation cycles in the CCI--$L_{\rm X}$ plane remains difficult in simulations because gas concentration measurements are limited by the $\sim$35\,Myr full-snapshot cadence. Among the four simulated galaxies, \simb~exhibits the clearest long-lived event for which a coherent trajectory can be followed over the final Gyr.

Figure~\ref{fig:timetrack_CCI_vs_Lx} shows the evolutionary path of \simb\ in the $\Sigma_{R_{32}}/\Sigma_{R_{256}}$--$L_{\rm X}^{\rm AD}$ plane, where $L_{\rm X}^{\rm AD}$ is computed at high temporal resolution from BH-specific outputs (every $\lesssim$0.1\,Myr at late times\footnote{At earlier times the cadence varies between 0.01 and 0.8\,Myr, while in the last Gyr it ranges from 0.01 to 0.1\,Myr.}), whereas gas concentration is tied to the much sparser snapshot spacing. For each timestamp in $L_{\rm X}^{\rm AD}$, we assign the gas concentration from the nearest snapshot, preserving the rapid variability in $L_{\rm X}^{\rm AD}$ but imprinting discretization in gas concentration.\footnote{As a result, several consecutive BH outputs may share the same gas concentration value, producing horizontal bands in Figure~\ref{fig:timetrack_CCI_vs_Lx}.} 
The corresponding time series for all four simulations are shown in Figure~\ref{fig:all_CCI_timeplots}, making explicit that the dense temporal sampling comes from the BH-specific luminosity outputs, while the concentration evolution is limited by the snapshot cadence and can differ between the fixed-aperture and dynamically defined measurements.
Because the observational branches from \citet{Garcia-Burillo2024} are defined using the fixed $\Sigma_{\rm 50\,pc}/\Sigma_{\rm 200\,pc}$ concentration, while $\Sigma_{R_{32}}/\Sigma_{R_{256}}$ uses adaptive apertures that vary in time, we do not interpret the branch locations as a quantitative classification for the simulated points. Instead, the observational branches are shown as a qualitative reference for the sense of the proposed evolutionary sequence. The adaptive concentration is used here to trace whether the simulated galaxy follows a similar relative loop in concentration--$L_{\rm X}$ space when the nuclear gas distribution is measured on the resolved BH-kernel scale.

We overplot \simb's trajectory by applying cubic spline interpolation and sampling the result at 500 uniformly spaced times, where the line colour encodes time evolution (from white to black) over the 1\,Gyr period.
In this case, \simb\ spent $\sim$200\,Myr in the high-concentration, high-$L_{\rm X}$ region of the diagram, qualitatively overlapping the part of the observational plane associated with the AGN feedback branch in GATOS (red line). It then experienced a strong blowout phase into the high-luminosity/low-concentration part of the diagram, with AGN winds clearing out gas on $\sim$300–1000\,pc scales (Figure~\ref{fig:all_overview}; top panel), before subsequently moving to the low-luminosity/low-concentration quadrant as the nuclear gas reservoir remained depleted and the AGN faded.
Over the next $\sim$600\,Myr, galaxy \simb\ slowly recovered its nuclear gas reservoir while approaching the AGN build-up branch of GATOS (blue line).
This rare, long-lived depletion event produces a coherent trajectory even at the coarse simulation snapshot cadence, supporting the picture of AGN fuelling and feedback self-regulation cycles leaving an imprint in the concentration--$L_{\rm X}$ relation \citep{Garcia-Burillo2021,Garcia-Burillo2024}. A finer snapshot cadence or on-the-fly estimates of nuclear gas concentration could reveal the more common $\sim$10--100\,Myr self-regulation loops and enable a more direct comparison to the plausible evolutionary tracks inferred from observations.
However, we cannot uniquely attribute either the cavity or the drop in concentration to AGN feedback alone. Simulations without BHs show that stellar feedback can also produce central cavities and time-variable structures in the inner few hundred parsecs \citep{Torrey2017}, including CMZ-like rings and gaps \citep[e.g.][]{Orr2021}. 
This degeneracy between stellar and AGN-driven effects is an important caveat when using changes in concentration or cavity size to infer the causal impact of AGN feedback on the nuclear gas reservoir.


Once AGN feedback has excavated a central cavity, the timescale for re-fuelling depends on the size of the cavity and how efficiently angular momentum is transported through the galaxy. A range of processes can contribute, from galaxy interactions to large-scale bars, spiral arms, nuclear asymmetries, and clumps \citep[e.g.][]{Kormendy2004,Hopkins&Quataert2010,Silverman2011,Storchi-Bergmann2019,Audibert2021,Combes2023,Nagele2023,Kataria2024}, with recent simulations also highlighting the role of wind-driven pileup and reaccretion events in rebuilding central gas reservoirs \citep{Mercedes-Feliz2025_pileup}. Recent analysis of FIRE discs shows that torques from gravity, hydrodynamical forces, and SN remnants all play a role transporting gas across our Milky Way-mass galaxies \citep{Trapp2024}. In the inner region, gravitational torques from stars dominate the removal of angular momentum \citep[as in luminous AGN at higher redshift;][]{Angles-Alcazar2021,Hopkins2024_zoom1}, while feedback and gas–gas interactions redistribute it within the disc and its immediate surroundings. Large-scale stellar bars, when present, are more likely to form and longer-lived in kinematically cold discs, where they can persist for many rotation periods (up to $\sim$100 bar rotations) and affect gas inflows over long timescales \citep{Rosas-Guevara2022,Rosas-Guevara2025,Ansar2025,Kalita2026}. These timescales are generally longer than the individual AGN fuelling and feedback cycles studied here, which may help explain why large statistical samples often find only weak or no enhancement of average AGN activity in barred galaxies compared to unbarred systems \citep[e.g.][]{Cisternas2015,Galloway2015,Silva-Lima2022,Zee2023,LaMarca2026}.

In addition to the high time variability challenge, the treatment of BH accretion and feedback introduces additional sources of uncertainty. In the FIRE-3 simulations analyzed here, the subgrid accretion disc adopts fixed prescriptions for how inflowing gas drives BH accretion as well as the associated radiative and mechanical efficiencies. In reality, the accretion disc timescale, the radiative output, and the relative importance of winds and jets are expected to depend on the state of the accretion flow, transitioning between radiatively inefficient hot flows, radiatively efficient thin discs, and slim-disc regimes as a function of Eddington ratio and accretion geometry \citep[e.g.][]{Yuan&Narayan2014,Sadowski2015,Ryan2017}, or following the scalings of recently proposed magnetically-dominated AGN discs \citep{Hopkins2024_zoom2,Kaaz2025}. 
In post-processing experiments, allowing the radiative efficiency $\epsilon_{\rm rad}$ to decrease at low Eddington ratios ($\lambda_{\rm Edd} < 0.03$), as expected for radiatively inefficient discs \citep[e.g.][]{Yuan&Narayan2014,Koudmani2024,Rennehan2024}, broadens the predicted $L_{\rm X}^{\rm BH}$ distribution towards lower luminosities, in better agreement with observations but still insufficient to reproduce the concentration--$L_{\rm X}$ relation in the GATOS survey. 
Recent subgrid feedback models in galaxy-scale and cosmological simulations are beginning to incorporate such disc state-dependent efficiencies more explicitly \citep[e.g.][]{Koudmani2024,Rennehan2024,Beckmann2025,Husko2026}, and GRMHD-based multi-zone simulations are enabling prescriptions for BH accretion and feedback efficiencies that can further help bridge scales in simulations \citep[e.g.][]{Cho2024,Cho2026,Guo2025,Su2025}.
Implementing similar prescriptions in the FIRE-3 framework could modify the timing and amplitude of AGN self-regulation events.

The detailed properties of the nuclear cavities may also depend on the AGN feedback coupling implementation. In the present model, AGN feedback is not deposited through kernel-weighted heating or momentum kicks to selected pre-existing gas elements within the BH kernel \citep[e.g.][]{Schaye2015,Angles-Alcazar2017a,Weinberger2017,Dave2019}, which can artificially enhance coupling to the cold ISM gas component and imprint nonphysical feedback scales.  Instead, new hyper-refined wind particles are launched from the immediate vicinity of the BH and subsequently couple hydrodynamically to the surrounding multiphase ISM, more similar to high-resolution AGN wind injection in the moving-mesh {\sc BOLA} model \citep{Costa2020,Ward2024}. The resulting cavity radius is therefore a prediction of the model and not prescribed  by the size of the BH kernel ($R_{256}$). Nevertheless, quantitative cavity properties may still depend on numerical and subgrid choices, including the wind velocity and mass loading, launch geometry, radiative coupling, and the subsequent interaction of the outflow with the multiphase ISM. Because we do not vary these feedback prescriptions at fixed initial conditions in the present study, we cannot determine how invariant the detailed cavity sizes, depths, morphologies, or lifetimes are to such modelling choices. While there is encouraging similarity in simulated and observed cavity sizes and morphologies, we therefore interpret the occurrence and qualitative evolution of depleted nuclear states more robustly than their precise quantitative cavity properties.

Future work should push toward simulations that combine the cosmological environments modeled here with the sub-pc resolution needed to follow accretion flows onto the BH and to resolve the structure of the accretion disc. Lagrangian hyper-refinement techniques applied to massive galaxies at $z\sim 2$ have already demonstrated that it is possible to track gas transport from halo scales down to $\sim$0.1\,pc \citep{Angles-Alcazar2021}. This increased dynamic range is crucial to capture explicitly the time delay between increasing gas concentration on $\sim$50--200\,pc scales and the onset of AGN activity, which may enable simulations to better reproduce observed Seyfert galaxies in the high-concentration--low-luminosity regime. The FORGE'd in FIRE simulations extend hyper-refinement techniques to radiation-MHD models that connect galaxy scales all the way down to the BH accretion disc, resolving the suppression of star formation on sub-pc scales and the structure of the inner accretion flow \citep{Hopkins2024_zoom1,Hopkins2024_zoom2,Hopkins2025_zoom}. 
The emerging magnetically-dominated accretion disc is characterized by significantly shorter inflow timescales than traditional $\alpha$-discs, increasing the frequency of high-luminosity AGN phases and potentially strengthening their association with central cavities, in better agreement with late-type Seyfert observations. This suggests that a more state-dependent accretion disc response time could help improve agreement with the GATOS feedback branch. By reducing the delay between gas reaching sub-pc scales, AGN brightening, and feedback-driven clearing on $\sim$50--200 pc scales, such models may increase the fraction of snapshots in which high $L_{\rm X}$ coincides with recently evacuated nuclear cavities.

\section{Conclusions} \label{sec:conclusion}
In this study, we used high-resolution cosmological zoom-in simulations implementing FIRE-3 physics \citep{Hopkins2023_fire3,Byrne2024} to investigate the connection between BH accretion, AGN feedback, and the spatial distribution of nuclear gas in Milky Way-mass galaxies. By computing gas concentration indices analogous to observational metrics from the GATOS survey \citep{Alonso-Herrero2021,Garcia-Burillo2021}, we have systematically explored whether the bimodal concentration–X-ray luminosity trends reported for nearby Seyfert galaxies arise naturally in a fully cosmological setting.
Our key findings can be summarised as follows:
\begin{enumerate}[wide, labelwidth=!,itemindent=!, label=(\alph*)]
    \item All four simulated galaxies exhibit recurrent cycles of gas inflow toward the subgrid accretion disc, followed by increased BH accretion, AGN feedback-driven cavity formation, and a decrease in the inflow rate until the next fuelling event (Figure~\ref{fig:all_timeplots}). These cycles appear as recurrent loop-like trajectories in the $\dot{M}_{\rm AD}$--$\dot{M}_{\rm BH}$ plane (Figure~\ref{fig:m12f_MandM}), illustrating the self-regulatory nature of BH growth. Periods of high $\dot{M}_{\rm AD}$ coincide with high nuclear gas density and the subsequent increase in $\dot{M}_{\rm BH}$, while feedback-driven clearing of the nuclear region reduces $\dot{M}_{\rm AD}$ and subsequently $\dot{M}_{\rm BH}$, producing episodic behaviour on characteristic timescales $\sim$10--100\,Myr.
    
    \item The central gas distribution responds dynamically to AGN activity. Temporal fluctuations in gas concentration, whether measured using fixed physical apertures (50--200\,pc) or adaptive kernel radii ($R_{32}$--$R_{256}$), track the formation and dispersal of nuclear gas reservoirs (Figure~\ref{fig:m12i_multipanel}). Our simulations demonstrate that cavities in the nuclear gas can form even in the presence of ongoing inflows while AGN winds escape preferentially along the polar direction, highlighting the complex interplay between fuelling and feedback.

    \item Simulated galaxies span a range of nuclear gas concentrations and AGN X-ray luminosities broadly similar to those observed in GATOS and WISDOM, which is a significant success for cosmological simulations. However, they do not reproduce the  anti-correlation between gas concentration and $L_{\rm X}$ seen in GATOS, which provides a cleaner comparison sample for late-type disc galaxies. This difference suggests a meaningful tension: the simulations form similar feedback-driven cavities, but high-$L_{\rm X}$ phases are not preferentially associated with low nuclear gas concentrations as often as inferred from GATOS.

    \item Median radial gas surface density profiles indicate that higher accretion rates onto the subgrid disc generally coincide with steeper nuclear density profiles, reflecting efficient gas inflow (Figure~\ref{fig:LXbins}). In contrast, instantaneous BH accretion rates alone can be misleading indicators of nuclear gas content due to the buffering effect of the accretion disc reservoir. This suggests that the characteristic delay between nuclear gas inflow, subgrid BH accretion, and feedback-driven gas clearing is one of the main quantities constrained by the GATOS comparison. Higher X-ray luminosity in simulated galaxies can correlate or anti-correlate with central gas concentration in fuelling- versus feedback-dominated phases, while the GATOS survey indicates that the anti-correlation is more prevalent in luminous Seyfert galaxies.
    
\end{enumerate}

Taken together, our results indicate that spatially resolved molecular observations constrain not only whether AGN feedback can create nuclear cavities, but also when those cavities appear relative to AGN brightening. The FIRE-3 simulations produce recurrent self-regulation cycles and cavities on the relevant $\sim$\,50--200 pc scales, but the simulated timing does not yield the same prevalence of high-$L_{\rm X}$, low-concentration systems seen in GATOS. This points to the coupling between nuclear gas inflow, unresolved accretion-disc evolution, and wind feedback as the key model uncertainty. In particular, the accretion-disc depletion time and the coupling efficiency of AGN winds in the nuclear region may determine whether simulations populate a GATOS-like feedback branch or instead show the more stochastic behaviour seen here.

Future work should extend this analysis to larger samples of simulated galaxies across a broader range of masses and AGN activity levels, implementing hyper-refinement techniques to resolve fuelling and feedback explicitly at sub-pc resolution \citep[e.g.,][]{Angles-Alcazar2021,Mercedes-Feliz2023,Hopkins2024_zoom1}. Additionally, incorporating synthetic atomic and molecular line observations from the simulations \citep[e.g.,][]{Richings2022,Roy2026} will enable more direct comparison to observations, further clarifying the connection between nuclear gas distribution and AGN activity in the context of galaxy evolution.

\section*{Acknowledgements}
We thank the referee, Filip Hu{\v s}ko, for constructive comments that helped improve the paper.
The simulations were run using the Advanced Cyberinfrastructure Coordination Ecosystem: Services \& Support (ACCESS), supported by NSF grant ACI-1548562, including allocations TG-AST140023, PHY260171, and PHYS260190; Frontera allocations AST21010 and AST20016, supported by the NSF and TACC; Pleiades, via the NASA HEC program through the NAS Division at Ames Research Center.
JMF was supported in part by a NASA CT Space Grant.
DAA acknowledges support from NSF CAREER award AST-2442788, NASA grant ATP23-0156, STScI JWST grants AR-04357.001-A and AR-05366.005-A, an Alfred P. Sloan Research Fellowship, and Cottrell Scholar Award CS-CSA-2023-028 by the Research Corporation for Science Advancement. 
SGB acknowledges support from the Spanish grant PID2022-138560NB-I00, funded by MCIN/AEI/10.13039/501100011033/FEDER, EU.
CRA acknowledges support from the Agencia Estatal de Investigaci\'on of the Ministerio de Ciencia, Innovaci\'on y Universidades (MCIU/AEI) under the grant ``Tracking active galactic nuclei feedback from parsec to kiloparsec scales'', with reference PID2022$-$141105NB$-$I00 and the European Regional Development Fund (ERDF).
CAFG was supported by NSF through grants AST-2108230 and AST-2307327; by NASA throgh grants 80NSSC22k0809, 80NSSC22K1124, and 80NSSC24K1224; by STScI through grant JWST-AR-03252.001-A; and by BSF through grant \#202462.
MPS acknowledges support under grants RYC2021-033094-I, CNS2023-145506 and PID2023-146667NB-I00 funded by MCIN/AEI/10.13039/501100011033 and the European Union NextGenerationEU/PRTR.

\section*{Data Availability}
The data supporting the plots within this article are available on reasonable request to the corresponding author. FIRE-2 simulations are publicly available \citep{Wetzel2023,Wetzel2025} at \url{http://flathub.flatironinstitute.org/fire}. Additional FIRE simulation data, including initial conditions and derived data products, are available at \url{https://fire.northwestern.edu/data/}. A public version of the GIZMO code is available at \url{http://www.tapir.caltech.edu/~phopkins/Site/GIZMO.html}.



\bibliographystyle{mnras}
\bibliography{main} 




\appendix

\section{Time evolution of gas concentration--$L_{\rm X}$ relation}

\begin{figure*}
\includegraphics[width = 0.8\textwidth]{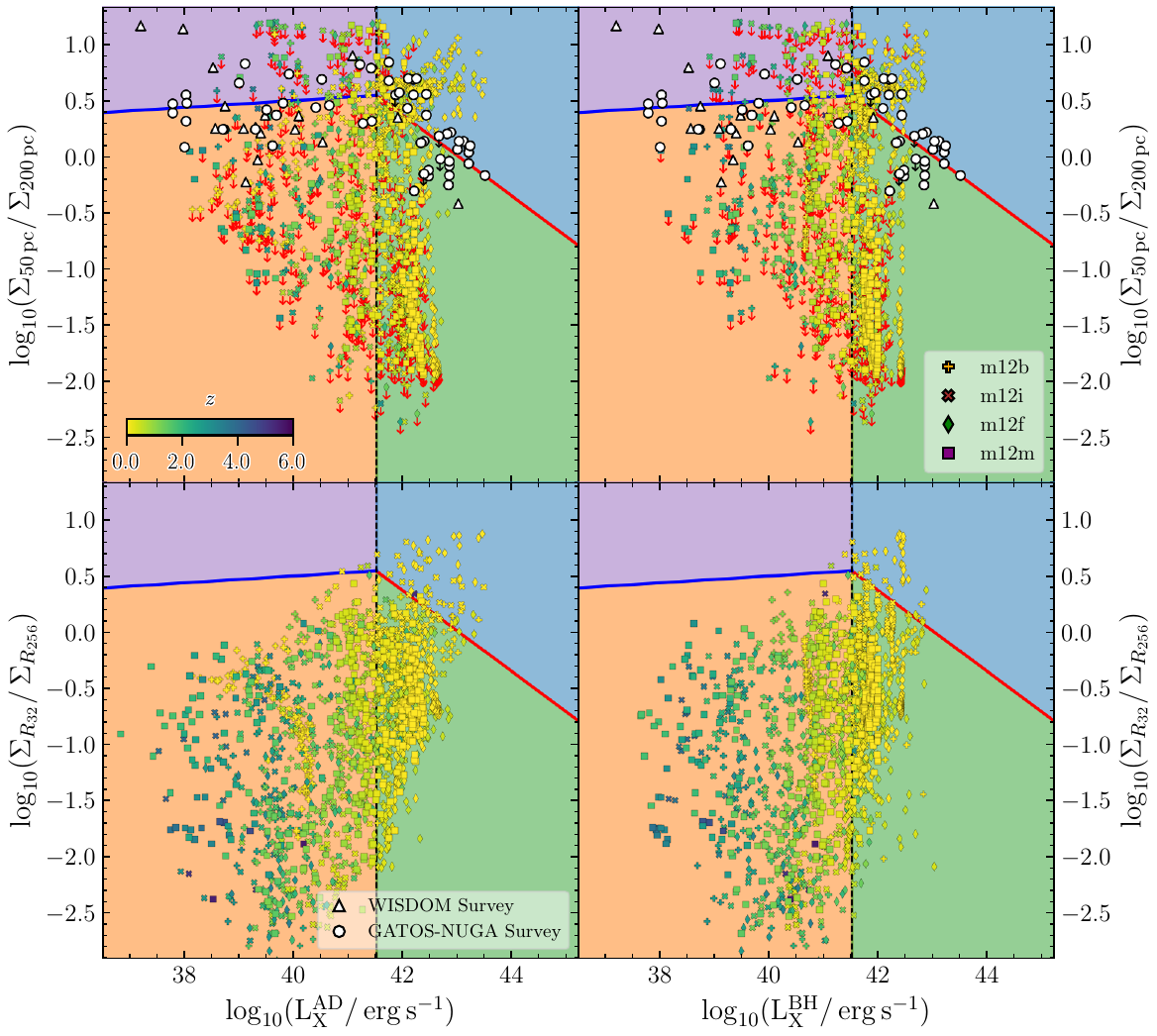}
\caption{Same as Figure~\ref{fig:comparison_to_obs}, but including all available times in the simulation. Simulated galaxies are denoted by squares, diamonds, triangles, and crosses, colour-coded by redshift ($z$), with higher redshift shown in purple and lower redshift in yellow. Upper limits are indicated by the red arrows for the top panels. Observations from the WISDOM \citep{Elford2024} and GATOS-NUGA \citep{Garcia-Burillo2024} surveys are indicated in white triangles and circles, respectively. Simulated galaxies cover a broadly similar range of parameter space as observations, but tend to drive larger cavities and reach lower gas concentrations than observations for a given X-ray luminosity.
}
\label{fig:comparison_to_obs_alltimes} 
\end{figure*}
Figure~\ref{fig:comparison_to_obs_alltimes} shows the gas concentration as a function of X-ray luminosity, similar to Figure~\ref{fig:comparison_to_obs} but for all available snapshots across our four simulations. The points are coloured according to redshift, higher redshift shown in purple and lower redshifts shown in yellow. 
By including all simulation snapshots, we are able to populate the low-luminosity region of the diagram, which provides a more complete evolutionary picture. 
The colour gradient, which indicates redshift, reveals a clear trend across all four galaxies: as time progresses and BHs grow (from high to low redshift), AGN generally become more luminous in these Milky Way-mass hosts. A notable exception is simulation $\simb$, which shows a decline in luminosity during its final Gyr (as discussed in \S\ref{sec:correlation}). 
In the top panels, the better sampling of low-luminosity systems increases the overlap with the GATOS and WISDOM surveys at ${\rm log}_{10}(\Sigma_{\rm 50\,pc}/\Sigma_{\rm 200\,pc}) > 0$ and $L_{\rm X} < 10^{41.5}\,{\rm erg}\,{\rm s}^{-1}$.
The bottom panels show that simulated BHs can reach luminosities $L_{\rm X} < 10^{38}\,{\rm erg}\,{\rm s}^{-1}$ with ${\rm log}_{10}(\Sigma_{R_{32}}/\Sigma_{R_{256}}) < 0$ at early times, with those low-luminosity data points absent in the top panels.
This can be explained by the presence of large cavities in the nuclear region, which are often larger than 200\,pc and therefore do not allow computing gas concentration based on the fixed physical apertures at 50--200\,pc.
The lack of a strong, universal correlation between gas concentration and X-ray luminosity suggests that the underlying relationship is highly dynamic and not easily captured, as discussed in \S\ref{sec:discussion}.

\begin{figure*}
\centering
\centerline{\includegraphics[width = 0.5\textwidth]{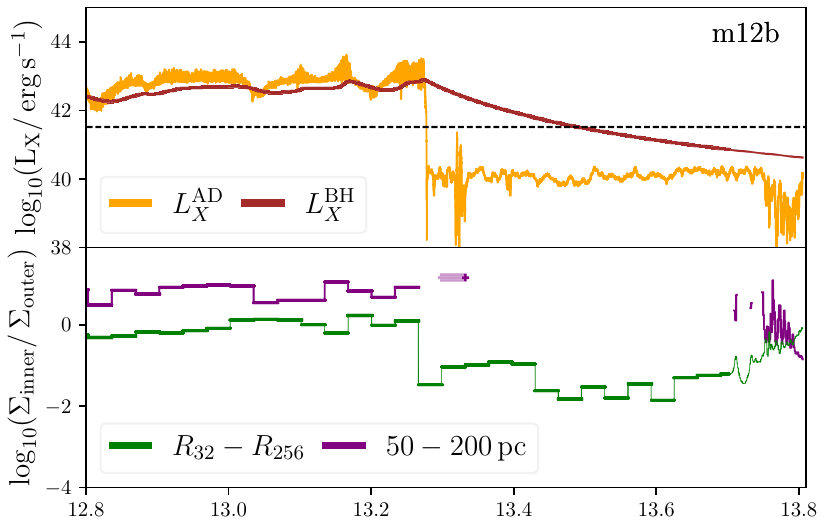}
\includegraphics[width = 0.5\textwidth]{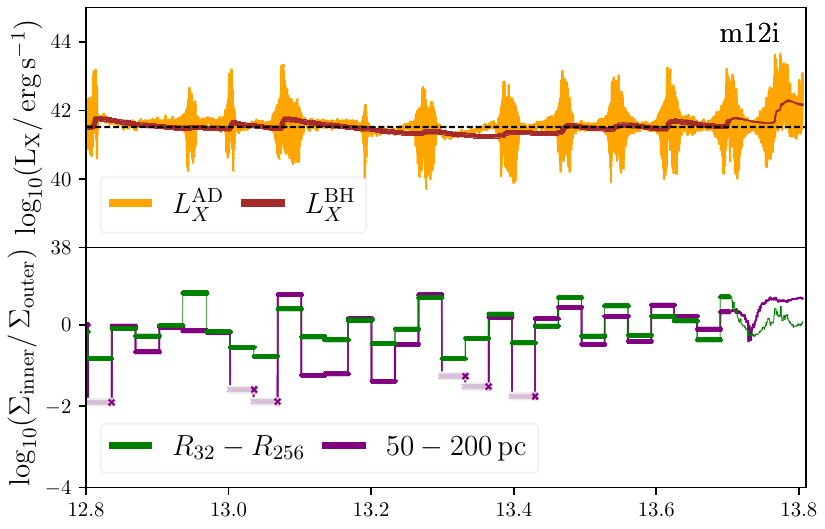}}
\centerline{\includegraphics[width = 0.5\textwidth]{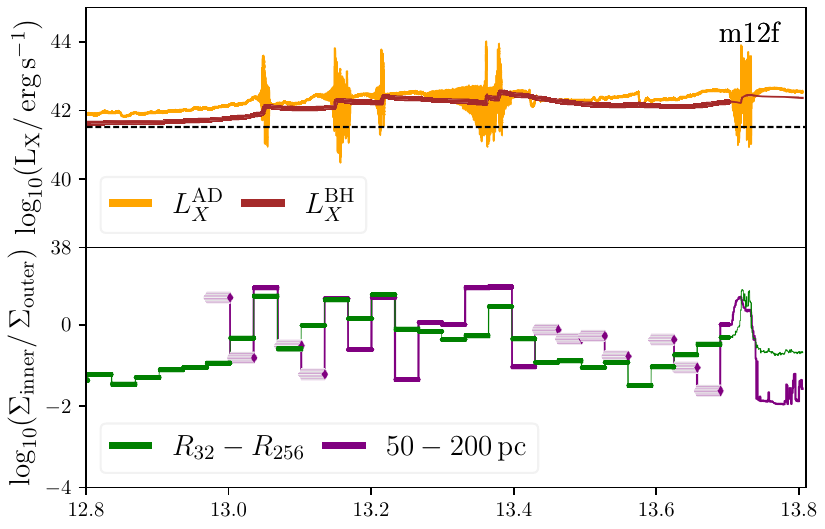}
\includegraphics[width = 0.5\textwidth]{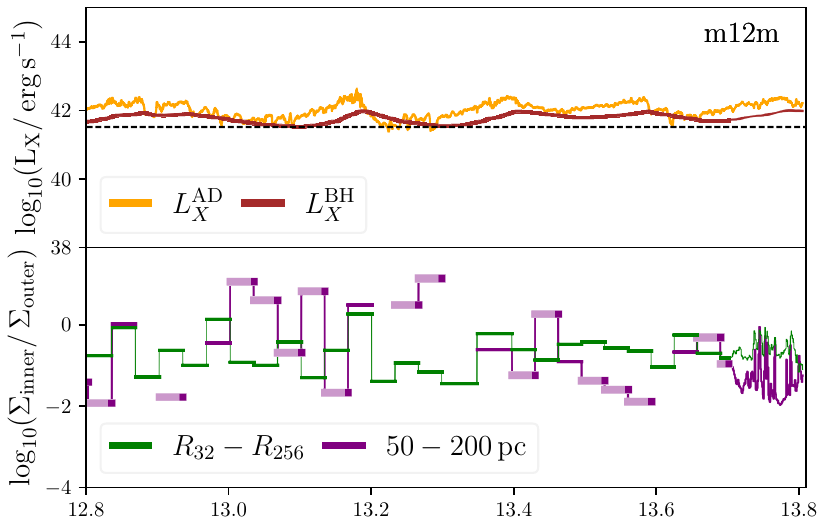}}
\vspace*{-2mm}
\caption{
Time evolution of the X-ray luminosity ($L_{\rm X}$; top panel) and nuclear gas concentration  ($\Sigma_{\rm inner}/\Sigma_{\rm outer}$; bottom panel) over the final 1\,Gyr for the four simulations. For each simulation, the top panel shows the luminosity inferred from the inflow rate onto the accretion disc, $L_{\rm X}^{\rm AD}$, and from the BH accretion rate, $L_{\rm X}^{\rm BH}$. The black dashed horizontal line marks the inferred transition luminosity from \citet{Garcia-Burillo2024}. The bottom panel shows the gas concentration measured using both the dynamically defined apertures, $R_{32}$--$R_{256}$, and the fixed physical apertures, 50--200\,pc. For the fixed-aperture measurement, outlined points indicate upper limits. The high-cadence BH-specific outputs provide the $L_{\rm X}$ time series, while each output is assigned the concentration measured from the nearest available simulation snapshot. Both the amplitude and timing of nuclear gas concentration changes vary across the simulations, and the fixed-aperture and dynamically defined measurements do not always evolve identically.
}
\label{fig:all_CCI_timeplots} 
\end{figure*}

Figure~\ref{fig:all_CCI_timeplots} shows the time evolution of the X-ray luminosity and nuclear gas concentration for the four simulations over the final 1\,Gyr. This provides the time-domain view corresponding to the concentration--luminosity diagrams shown in Figure~\ref{fig:comparison_to_obs} but expanded using the high-cadence BH-specific data outputs as done with Figure~\ref{fig:timetrack_CCI_vs_Lx}. The BH-specific outputs provide high-cadence measurements of both $L_{\rm X}^{\rm AD}$ and $L_{\rm X}^{\rm BH}$, while the gas concentrations are computed from the full simulation snapshots. We therefore assign each BH-specific output the concentration measured from the nearest available snapshot. This preserves the short-timescale variability in $L_{\rm X}$, but produces step-like behaviour in $\Sigma_{\rm inner}/\Sigma_{\rm outer}$ because multiple BH outputs can be associated with the same snapshot-based gas measurement. We show two concentration definitions. The first uses the dynamically defined apertures $R_{32}$ and $R_{256}$, which enclose the nearest 32 and 256 gas elements around the BH, respectively. The second uses fixed physical apertures, comparing the gas surface density within 50\,pc to that within 200\,pc. These two definitions provide complementary measures of the nuclear gas distribution: the fixed-aperture ratio is closer to the observationally motivated concentration used in the main text, while the $R_{32}$--$R_{256}$ ratio follows the local gas distribution around the BH with apertures that vary according to the resolved gas sampling.

For the fixed 50--200\,pc concentration, outlined points indicate upper limits. These occur when no gas particle is present within the inner 50\,pc aperture. In these cases, we assign one gas-particle mass to the inner aperture when computing $\Sigma_{50\,{\rm pc}}/\Sigma_{200\,{\rm pc}}$, so the plotted value should be interpreted as an upper limit on the true fixed-aperture concentration. These points therefore mark times when the fixed inner aperture is depleted at the available mass resolution. The four simulations show diverse late-time behaviour in both luminosity and gas concentration. In some cases, the dynamically defined and fixed-aperture concentrations evolve similarly, while in others they differ in amplitude or timing. This difference is important for interpreting the corresponding concentration--luminosity phase-space tracks, since changes in $L_{\rm X}$ are sampled at much higher cadence than changes in the gas distribution, and the two concentration definitions do not always trace identical nuclear gas evolution.


\bsp	
\label{lastpage}
\end{document}